# Measuring Dielectric Properties and Surface Resistance of Microwave PCBs in the K-band

Victor N. Egorov[1)], Vladimir L. Masalov[1)], Yuri A. Nefyodov[2)], Artem F. Shevchun[2)],
Mikhail R. Trunin[2)], Victor Zhitomirsky[3)], Mick McLean[3)]

[1)]East-Siberian Research Institute of Physico-Technical and Radioengineering Measurements,
57, Borodina str., Irkutsk, 664056, Russia

[2)]Institute of Solid State Physics RAS, Chernogolovka, Moscow district, 142432, Russia

[3)]Scientific Generics, Harston Mill, Harston, Cambridge, GB2 SGG, UK

**ABSTRACT**

The theoretical model is fully developed and the test rig is designed for the measurements of microwave parameters of unclad and laminated dielectric substrates. The geometry of the electromagnetic field in the resonator allows dielectric measurements with electric field component orthogonal to the sample surface. The test rig was completely automatized for measurements of the following parameters: (i) dielectric constant ($\varepsilon$) of the dielectric substrate in the range from 2 to 10, (ii) loss tangent (tanδ) of the dielectric substrate in the range from $10^{-4}$ to $10^{-2}$, and (iii) microwave ohmic loss at the interface between the metal layer and the dielectric material in the range from 0.01 Ω to 0.2 Ω. Measurements for a number of commonly-used microwave PCB materials were performed in the frequency range from 30 to 40 GHz and over a temperature range from -50 $^0$C to +70 $^0$C.

## I. INTRODUCTION

It is well-known that until comparatively recently microwave and millimetre wave technologies have been predominantly exploited in military applications because of the high cost of componentry. Due to advances in telecommunications over the last decade, however, the cost of components operating at frequencies around 5 GHz and below has been dramatically reduced. Industry observers predict that the same dramatic reduction in component cost will happen in the frequency band below 100 GHz over the next ten years as a result of developments in semiconductor high-frequency electronics, new compact antennas, and advances in integrated circuits (MMICs). This process will be catalysed by such emerging high-volume industrial applications as:

- 77 GHz standard for car anti-collision systems;
- 40 Gb/s optical telecom equipment,
- "millimetre wave" high-speed point-to-point wireless links for local and metropolitan networks and broadband Internet access (30- and 40-GHz bands in Europe and 70-, 80-, and 90-GHz bands in the USA).

Essential parameters needed for the efficient design of integrated microwave circuits are dielectric properties, the degree of passive inter-modulation, and the microwave copper resistance of the PCB substrate on which the active elements are mounted. As components are increasingly miniaturised and frequencies increased, the need for accurate low-loss dielectric measurements of substrate materials increases. Simultaneously thinner substrates with low permittivity are used for higher wiring density and smaller embedded passive circuit elements. Accurate dielectric measurement methods are needed for MMIC design both for the minimisation of crosstalk and the characterisation of propagation speed. The measurement of the properties of thin materials presents a particular challenge in that uncertainties in thickness of the specimen directly translate into uncertainties in dielectric parameters.

Microwave PCB materials are usually composite materials and may be anisotropic. The composites usually consist of a mixture of plastics and ceramics, together with reinforcing materials. The degree of anisotropy in permittivity for normal and in-plane orientations of the electric field could reach 5% for fibre PTFE/glass composites and might exceed 20% for Woven PTFE/Glass materials [1]. In any measurement technique the structure of the incident electromagnetic field should be similar to the operating field in a real microstrip or microwave printed circuit. Most PCBs operate with electric field primarily normal to the plane of the sheet. An incident electromagnetic field should therefore have an electric field component orthogonal to the sample surface.

The properties of copper laminated PCB materials are well understood at the frequencies below 10 GHz where they are widely used. The established microstrip test standard IPC-TM-650 can be used for dielectric measurements below 10 GHz. The test uses an imaged resonator pattern on an actual PCB substrate [2]. The resonance frequency, bandwidth and electrical length are used to determine the dielectric parameters of the board material. However, the high radiation loss of the resonator pattern reduces the sensitivity of the test method. Microstrip-based tests do not allow the dielectric and ohmic losses to be measured separately. Microwave losses due to the copper surface resistance are taken into account by using table values for copper resistance, which are applicable only for an ideally flat copper surface. The non-reproducibility of the rig connection impedance at millimetre wave frequencies limits the accuracy of this method. Microstrip testing is also very time consuming even if dielectric parameters are measured at frequencies below 10 GHz. At higher frequencies this type of test becomes impractical because of the increasing difficulty of optimising the probe gap for microwave measurements.

Dielectric measurement methods at frequencies below and above 10 GHz can also be based on the use of a microwave bulk resonator. Resonant measurement methods represent the most accurate way of obtaining dielectric constant and loss tangent with unclad thin materials, - which are just inserted inside the cavity. The high value of the unloaded quality factor $Q_0$ of the bulk resonator enables measurements of the smallest losses in the test materials. Potentially simple and reproducible test methods based on bulk resonator measurements have not, however, been accepted as industrial measurement tools, since most resonant techniques yield permittivity in the plane of the sample instead of the value of permittivity orthogonal to the surface of a thin substrate.

Nevertheless different methods for making microwave measurements of in-plane dielectric parameters of thin materials based on bulk resonators have been developed in many laboratories and the results widely published [1,3-5]. The open hemispherical resonator [3,4,6] is one example of a very sensitive instrument for in-plane dielectric measurements of very low loss and flat specimens with diameters much greater than the wavelength. There are two problems with this approach: non-flatness of real samples and large resonator sizes, which limit the application of this technique for measurements in a wide temperature range. The cylindrical $H_{01p}$ cavity has been used by a few groups to measure both in-plane dielectric parameters for thin dielectric samples as well as surface resistance for ordinary metals and low- and high temperature superconductors [7-10]. However, in order to measure accurately the surface resistance of the test sample, a significant difference between the ohmic loss in the sample metallisation and in the walls of the resonator should be provided. In the case of copper laminated PCB samples such a difference might be attained only if the temperature of the metallic (typically copper) resonator is kept considerably below the sample temperature.

Measurement techniques which use dielectric split-cavity resonators are well established for measuring the parameters of thin dielectrics because of the inherently high values of their quality factors. For instance a cylindrical $H_{01\delta}$ dielectric split resonator [11] made from thermostable high permittivity ceramic has been successfully used for in-plane dielectric film measurements at frequencies below 10 GHz, but it was found unsuitable for measurements at higher frequencies due to increased loss tangent in ceramic materials. A full-model theory of split-cavity resonators including the effects of fringe fields at the gap in the cavity has been developed at NIST. Further details of this technique as well as of similar split-post dielectric resonator (SPDR) techniques for the measurement of in-plane properties of thin dielectric materials is available in NIST review papers [1].

Dielectric split-cavity resonators machined from sapphire can operate at much higher frequencies than those made from high permittivity ceramics. The sapphire disk "whispering gallery" (WG) resonator [12] has a $Q_0$ about 40000 at room temperature for frequencies in the range of 40 GHz (wavelength $\lambda$ of about 8 mm). The typical diameter of a split-cavity sapphire resonator is about $1.5\lambda$. Such resonators have been proposed for measuring free dielectric films [13] and have been successfully applied for the measurement of aluminium oxide film on aluminium substrates. There are two WG mode types: quasi-$E$ (or $HE$) and quasi-$H$ (or $EH$) with high a $Q_0$ for large azimuth mode index $n$>>1. They can be used for dielectric substrate measurements with orthogonal and tangential microwave $E$-fields respectively. In this paper we report on a successful project to measure the dielectric properties of thin films using the electric field component of an incident electromagnetic field which is orthogonal to the sample surface.

Our development of an automated test rig for the measurement of thin dielectric films was part of a programme of materials measurement research sponsored by the UK Department of Trade and Industry*. Our goal was to evaluate both theoretically and experimentally the uncertainties of the sapphire disk WG resonator technique for measurements of out-of-plane dielectric properties of thin materials. Our results support the case for establishing this method as a new unified standard for the 10 GHz to 100 GHz frequency range. An accuracy of 1 % for permittivity measurements of thin dielectric materials and a resolution of the order of $10^{-4}$ for their loss tangent has been shown at 40 GHz. Materials with the losses ranging from $10^{-2}$ to $10^{-4}$ have been successfully measured using up to five different resonance $HE_{n11}$ modes. Acceptable accuracy of measurement was provided both for extremely thin substrates with a thickness down to 30 µm as well as for very thick substrates with a thickness exceeding 1 mm. This became possible because the WG resonator achieves a substantial filling factor value even with very thin substrates. The high unloaded quality factor of the resonator also helps to measure accurately very small changes to the resonance frequency. To the best of our knowledge our project presents the first demonstration of a simple method for measuring the dielectric properties of thin materials when the electric field component is orthogonal to the sample surface. It is these orthogonal components are the most important for designers of microwave circuits.

The only current standard methodology which makes possible measurements of dielectric properties of thin materials with the electric field component orthogonal to the sample surface is based on microstrip testing. As discussed above the microstrip method is labour intensive, requires considerable skill to implement and does not scale well above 10 GHz.

In contrast our technique can be easily scaled up for frequencies above 10 GHz. Our technique is particularly useful for measurements of microwave PCB materials, as there is no need for complex machining and no surface structures need be imaged. Because of its fundamental simplicity our method can be used non-destructively by relatively unskilled staff either in the laboratory or on the assembly line.

Also for the first time our test method allows the measurement of the effective microwave surface resistance of laminated metal at the interface between the laminated material and the dielectric material. The experimental values of surface resistance are important both for modelling the properties of integrated circuits and for qualifying particular PCB materials for high power applications.

A significant and universal problem with making dielectric measurements with an orthogonal field is the so-called "residual air-gap" which exists due to the micro-roughness at the contact between the flat resonator and the specimen surfaces. As a result an effective "residual air-gap" should be usually taken into consideration in the electrodynamic model of the measured structure.

In the next chapter we discuss the structure of the electromagnetic field in the resonator and the theoretical model used for calculating the microwave parameters for unclad and laminated substrates. In Chapter 3 we present further details of the test rig design, the experimental procedures, and raw data processing. In Appendix A we present further details of the electromagnetic modelling of the sapphire disk "whispering gallery" resonator. Finally in Appendix B we give measurements results for a number of commonly-used microwave PCB materials in table form.

## II. ELECTRODYNAMICS

### *A. MEASUREMENT STRUCTURE BASIC*

Below we describe the resonance mode structure of a dielectric cylinder (Fig. 1a) with diameter $2a$ and height $L$, which is separated from a metallic plane by a dielectric layer of height $t$ and a gap of height $d$. If the component of the electric field in the $z$ direction $E_z(z)$ is an even function of $z$, then the plane $z = 0$ (metallic surface) behaves as a so called "electric wall" for which the following boundary conditions are satisfied: $E_{3r} = E_{3\varphi} = 0$. The electrodynamic structure of the modes in such a case is equivalent to the modes of the split dielectric resonator with a dielectric layer of double height $2t$ in the slot (Fig. 1b).

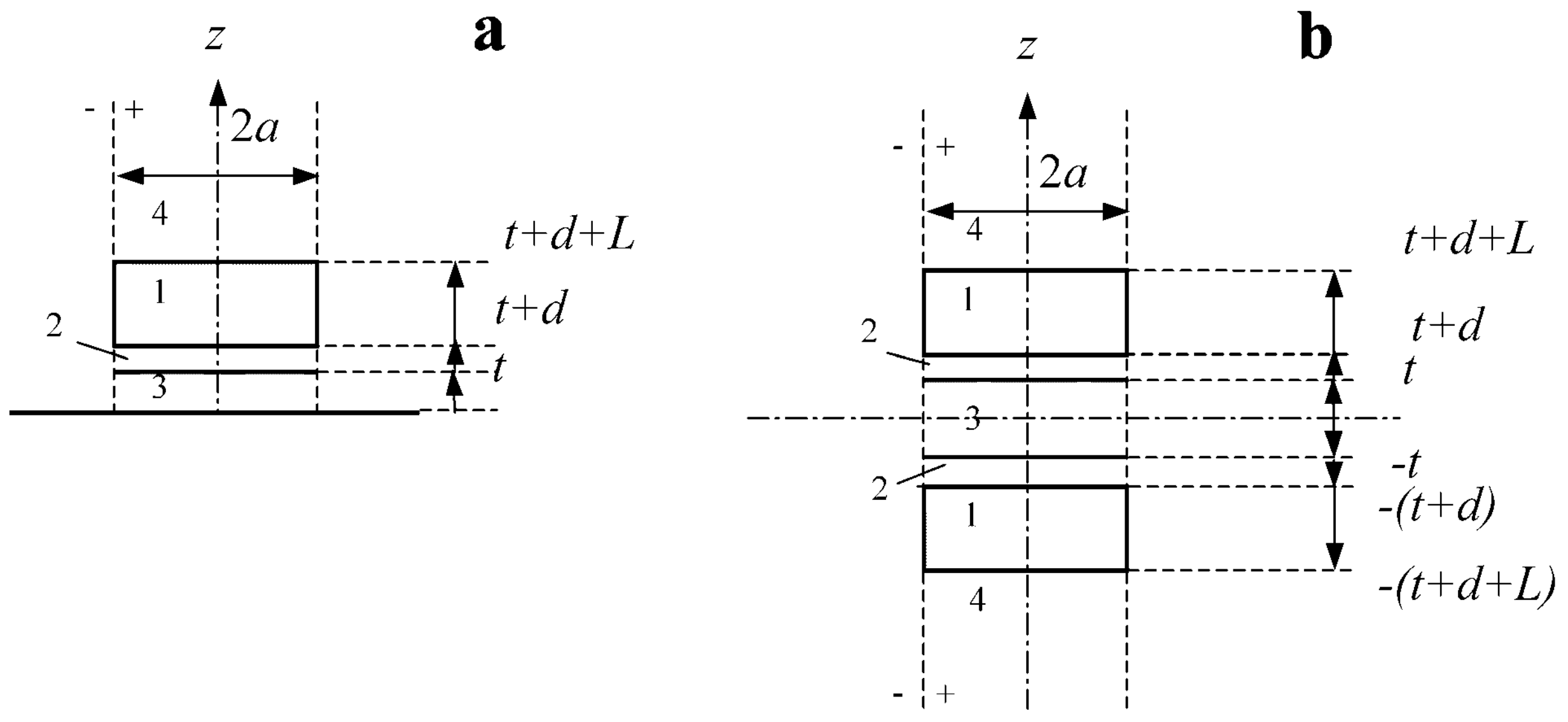

Fig.1. a) dielectric resonator (1) above a metallic plane with dielectric layer (3) inserted in between, the residual air gap (2) is also shown; b) dielectric layer in the split dielectric resonator.

The relative permittivity of a dielectric resonator is characterised by a tensor

$$\hat{\varepsilon}_1 = \begin{pmatrix} \varepsilon_{1\perp} & 0 & 0 \\ 0 & \varepsilon_{1\perp} & 0 \\ 0 & 0 & \varepsilon_{1\parallel} \end{pmatrix}$$

which determines its electric properties, and by a scalar $\mu_1$ for the magnetic properties. Symbols || and $\perp$ are related to the components of $\hat{\varepsilon}_1$ in the direction along the optical (geometrical) axis and in the plane perpendicular to this axis, respectively. We use $\varepsilon_3, \mu_3$ for the isotropic dielectric layer, and $\varepsilon_2, \mu_2$ for the ambient isotropic space, which includes both the top space (4) and the gap (2). We analyse the electromagnetic resonance modes by the method of approximate separation of variables with one-mode approximation of the fields at all fractional volumes of the resonator [6]. In this approach an electromagnetic field inside the resonator within the boundaries $t + d \le z \le t + d + L$ is represented in the form of linear combination of standing $E$ - and $H$ - waves, which forms a hybrid standing $HE$ or $EH$ wave along the $z$ - axis. Transverse (on $r, \varphi$ coordinates) field distribution in the gap (2), the dielectric layer (3) and the top space (4) is assumed the same as in the disk (1). Further details of the derivation of equations for the electromagnetic field components are given in Appendix A. The longitudinal wave numbers $h_{1E}$ and $h_{1H}$ of the $E$ - and $H$ -waves respectively are the same and are equal to the longitudinal wave number of the hybrid wave in the disk 1: $h_{1E} = h_{1H} = h$ .

The boundary conditions $E^{+}_{1\varphi} = E^{-}_{1\varphi}, H^{+}_{1\varphi} = H^{-}_{1\varphi}$ for inside (+) and outside (-) field components at $r = a$ within the limits $t + d \le z \le t + d + L$ $(i = 1)$ define the equation of a circular "dielectric post resonator" with single axis anisotropy [14]:

$$\left[\frac{\varepsilon_{\parallel} J_n'(\alpha\chi_1 a)}{\alpha\chi_1 a J_n(\alpha\chi_1 a)} - \frac{H_n^{(2)\prime}(\chi_2 a)}{\chi_2 a H_n^{(2)}(\chi_2 a)}\right]\left[\frac{\mu J_n'(\chi_1 a)}{\chi_1 a J_n(\chi_1 a)} - \frac{H_n^{(2)\prime}(\chi_2 a)}{\chi_2 a H_n^{(2)}(\chi_2 a)}\right] - \left(\frac{nh}{k_2}\left[(\chi_1 a)^{-2} - (\chi_2 a)^{-2}\right]\right)^2 = 0, \tag{1}$$

where $J_n(\chi_1 a), H_n^{(2)}(\chi_2 a)$, $J_n'(\chi_1 a), H_n^{(2)\prime}(\chi_2 a)$ are the Bessel and Hankel functions of the order $n$ and their derivations, $\chi_1, \chi_2$ are the inside (+) and outside (-) transverse wave numbers respectively, $\varepsilon_{\parallel} = \varepsilon_{1\parallel}/\varepsilon_2$, $\mu = \mu_1/\mu_2$, $\alpha = \sqrt{\varepsilon_{1\parallel}/\varepsilon_{1\perp}}$. For $\varepsilon_{\parallel} = \varepsilon_{\perp} \equiv \varepsilon_{1\perp}/\varepsilon_2$ this equation is reduced into the equation of an isotropic "dielectric post resonator".

For $HE_{nmp}$ modes with odd longitudinal index $p = 2q + 1$ $(q = 0, 1, 2, ...)$ the boundary conditions at $z = 0$, $z = t$, $z = t + d$, $z = t + d + L$ result in the characteristic equation [15]:

$$hL - \operatorname{atan}\frac{h_{2E}\eta_{1E}}{h} - \operatorname{atan}\left(\frac{h_{2E}\eta_{1E}}{h}\tanh\left(\operatorname{atanh}\left(\frac{h_{3E}}{h_{2E}\eta_{3E}}\tanh(h_{3E}t)\right) + h_{2E}d\right)\right) - (p-1)\pi = 0\ , \tag{2}$$

where $h_{iE}$ are the longitudinal wave numbers in regions $i = 2, 3, 4$ (Fig. 1), $h_{2E} = h_{4E}$:

$$h = \sqrt{k_1^2 - \chi_1^2} = \sqrt{k_2^2 - \chi_2^2}\ , \quad h_{2E} = \sqrt{(\alpha\chi_1)^2 - k_2^2}\ , \quad h_{3E} = \sqrt{(\alpha\chi_1)^2 - k_3^2}\ ,$$

$$k_1 = k_0\sqrt{\varepsilon_{1\perp}\mu_1}\ , \quad k_2 = k_0\sqrt{\varepsilon_2\mu_2}\ , \quad k_3 = k_0\sqrt{\varepsilon_3\mu_3}\ , \quad k_0 = \omega\sqrt{\varepsilon_0\mu_0}\ .$$

$$\eta_{1E} = \varepsilon_{1\perp}/\varepsilon_2\ , \ \eta_{3E} = \varepsilon_3/\varepsilon_2\ .$$

The set of Eqs. (1), (2) defines the values $h, \chi_1$ and $k_1 = \sqrt{h^2 + \chi_1^2}$, which depend on the relative dielectric sample permittivity $\varepsilon = \varepsilon_3/\varepsilon_2$. Eq. (1) does not explicitly depend on $\varepsilon$ and for determination of $\varepsilon$ one should solve the Eqs. (1) and (2) in series with the values of $k_1, k_2$ at the measured resonant frequencies.

The schematic fragment of electric and magnetic field structure for $HE_{n11}$ resonant mode is shown in Fig. 2a. Fig. 2b demonstrates the $E_z(r,\varphi)$ field relief and its level lines for a particular $HE_{911}$ mode.

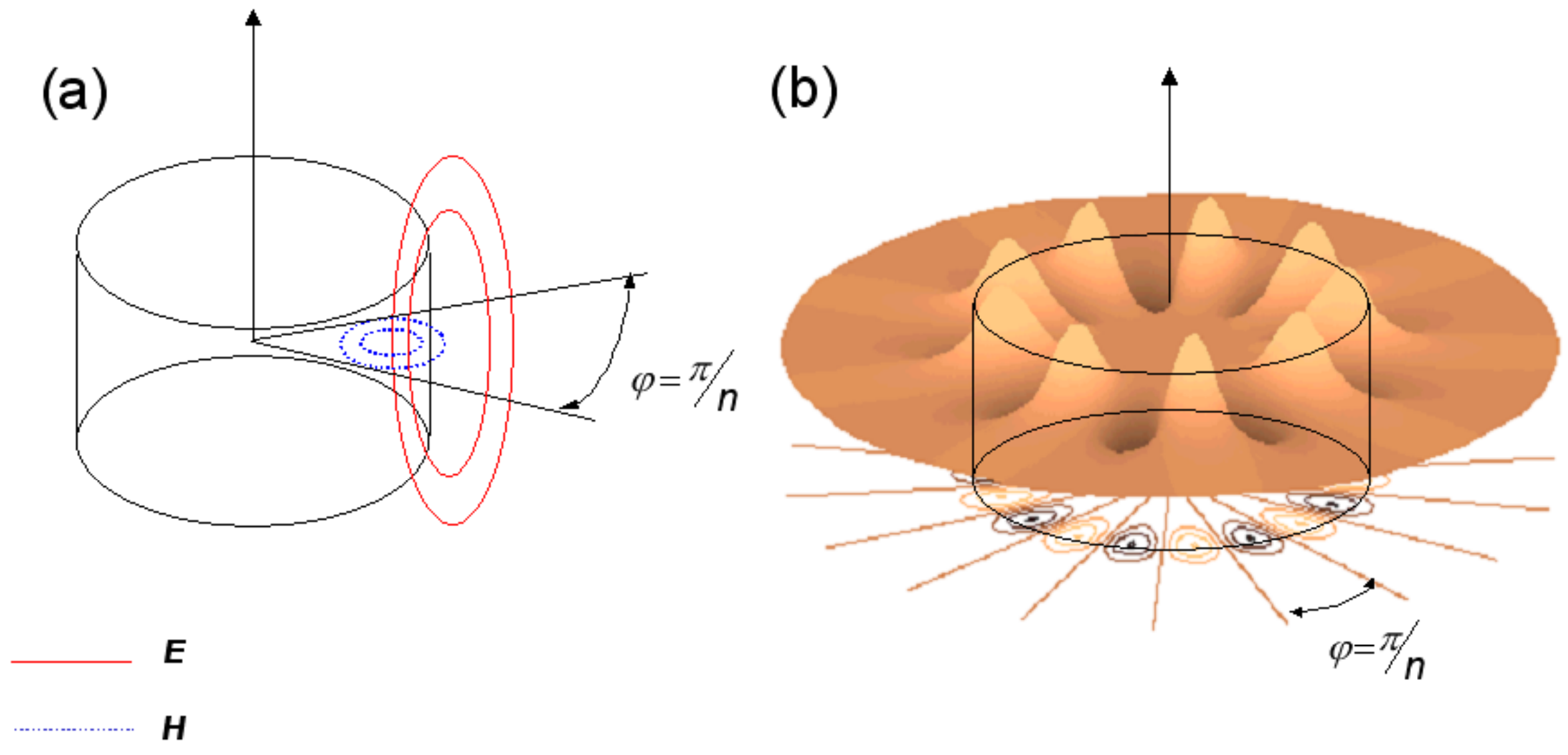

Fig. 2. (a) Schematic fragment of electromagnetic field structure for $HE_{n11}$ mode

(b) $E_z$ field relief and its level lines for n=9.

The electrodynamic model described by Eqs. (1), (2) does not take into account an influence of the part of the dielectric sample at $r > a$, $0 \le |z| \le t$, i.e. outside the resonator. If this sample volume is taken properly into account, the resonant frequencies will decrease and, hence, the models (1) and (2) - which do not take this into account - will overestimate values for $\varepsilon$. The dielectric sample volume outside the resonator is exposed to only a small part of the total electromagnetic energy. This enables one to correct the value of the dielectric constant by the perturbation method:

$$\varepsilon_{corr} = \varepsilon \cdot [1 - (\varepsilon - 1) K_{2E}] \, , \qquad (3)$$

where

$$K_{2E} = \frac{W_2^E}{W_\Sigma} = -2 \frac{\varepsilon_2}{\omega} \left( \frac{\partial \varepsilon_2}{\partial \omega} \right)^{-1}$$

and $W_\Sigma$ , $W_2^E$ are the total resonator energy and electric field energy outside the resonator disks. Factor $K_{2E}$ can be found by numerical differentiation of Eqs. (1), (2) with respect to the ambient media permittivity $\varepsilon_2$ [16]. $K_{2E}$ is typically of the order of 0.01 - 0.02. This value can be obtained more precisely by calibrating the test system using a range of dielectric samples with different thickness and known $\varepsilon$.

The frequencies of $HE_{nmp}$ modes in the resonator of height $L$ on the metallic plane ($t = 0$, $d = 0$) coincide with the frequencies of $HE_{nm2p-1}$ modes of the resonator with height $2L$ in free space. This becomes clear from Eq. (2) if one substitutes $L \to 2L$, $p \to 2p - 1$. This behaviour is rather obvious, because $HE_{nmp}$ modes with odd number $p$ have non-zero components $E_z, H_r, H_\varphi$ ("electric wall") in the symmetry plane of the single resonator.

### B. DIELECTRIC PERMITTIVITY AND LOSS TANGENT MEASUREMENTS OF NONMETALLIC SUBSTRATES

For measurements of the dielectric permittivity and loss tangent of the substrate, the foil is removed from both sides of the microwave PCB sample. The sample (substrate) is clamped between the plates of the split dielectric resonator (Fig. 1b and Fig. 7a below). In the experiment the values of the resonant frequencies of $HE_{n11}$ modes are determined. The basic data for calculating the dielectric permittivity of the sample using formulas (1), (2) are: (i) resonant frequency $f_n$ of $HE_{n11}$ modes with known azimuth index $n$, (ii) dimensions $a, L$ of dielectric resonator, (iii) sapphire dielectric permittivities $\varepsilon_{1\parallel}$, $\varepsilon_{1\perp}$, and (iiii) the thickness $t$ of the sample. The value of a "residual air-gap" $d$ is determined by the roughness of the surfaces of both the measured sample and the faces of the resonator and can not be measured directly. One can estimate this value from the condition that the measured value $\varepsilon_3$ should not dependent on the frequency of the measurements in a narrow frequency interval, which is defined by the frequencies of neighbouring (by azimuth index) $HE_{n-1,1,1}$, $HE_{n,1,1}$, $HE_{n+1,1,1}$ resonance modes. The frequency dispersion of low-loss dielectric samples in such a narrow frequency range is usually negligible in comparison with the uncertainty of the real measurements. The problem of the unknown residual air gap makes an additional contribution to inaccuracy, which slightly reduces the measured value $\varepsilon_3$. Uncertainty of the measurements depends on the frequency and is reduced with the increase of the azimuth index $n$ of the resonant mode.

The electromagnetic power $P_{\delta 3}$ directly absorbed by the sample (dielectric layer (3) in Fig. 1b) and electromagnetic energy $W_3^E$ stored in the sample layer are connected by the relationship

$$P_{\delta 3} = \omega W_3^E \tan\delta ,$$

where $\delta$ is the dielectric loss angle of a measured sample at a given electromagnetic field frequency $\omega$. In the ordinary approximation of the additive contribution of different losses, the power $P_{\delta 3}$ is connected with the total power loss $P_\Sigma$ and the unloaded quality factor $Q_{0\Sigma}$ of the resonator by the following equation

$$\frac{1}{Q_{0\Sigma}} = \frac{P_\Sigma}{\omega W_\Sigma} = \frac{P_{\delta 1} + P_{\delta 3} + P_{rad}}{\omega W_\Sigma} = \frac{1}{Q_{0DR}} + \frac{W_3^E}{W_\Sigma}\tan\delta + \frac{1}{Q_{rad}} , \qquad (4)$$

where $P_{\delta 1}$ is a dielectric loss power in the resonator dielectric disks only; $Q_{0DR}$ is a partial quality factor of these disks; $P_{rad}$ is the radiant loss; $Q_{rad}$ is the radiant quality factor of the resonator. It is easy to satisfy the condition

$$\frac{1}{Q_{0DR}} + \frac{W_3^E}{W_\Sigma}\tan\delta >> \frac{1}{Q_{rad}},$$

by choosing dimensions of the dielectric resonator. In this case we get from Eq. (4)

$$\tan\delta = K_{3E}^{-1}\left(\frac{1}{Q_{0\Sigma}} - \frac{1}{Q_{0DR}}\right), \tag{5}$$

where $K_{3E} = \frac{W_3^E}{W_\Sigma}$ is a filling factor of the resonator. The value of $K_{3E}$ can be obtained from the value of the stored energy by integrating the field components over the corresponding volumes of the resonator:

$$W_i^E = \frac{\varepsilon_0\varepsilon_i}{2}\int_{V_i}\left|\vec{E}\right|^2 dV, \quad W_\Sigma = \sum_i W_i^E = \sum_i W_i^H, \quad i = 1, 2, 3, 4 \tag{6}$$

where $W_i^{E,H}$ is the energy of electric or magnetic field stored in the $V_i$-volume of the resonator. Another way to calculate $K_{3E}$ is by numerical differentiation of the $\varepsilon_3(\omega)$ function obtained from Eqs. (1), (2) at measured resonant frequencies $\omega$ of the resonator with a sample inside [16]:

$$K_{3E} = -2\frac{\varepsilon_3}{\omega}\left(\frac{\partial\varepsilon_3}{\partial\omega}\right)^{-1}. \tag{7}$$

The quantity $Q_{0DR}$ in Eq. (5) is the unloaded quality factor of the resonator with a hypothetical sample, which has the dielectric permittivity of the real sample but has no loss ($\tan\delta = 0$). The value $Q_{0DR}$ is close to the unloaded $Q_{00}$ of the split resonator without a sample and can be found from the equation:

$$Q_{0DR}^{-1} = K_{1\parallel}\cdot\tan\delta_\parallel + K_{1\perp}\cdot\tan\delta_\perp, \tag{8}$$

where $K_{1\parallel} = \frac{W_{1\parallel}^E}{W_\Sigma}$, $K_{1\perp} = \frac{W_{1\perp}^E}{W_\Sigma}$ and $W_{1\parallel}^E + W_{1\perp}^E = W_1^E$. Here $W_{1\parallel}^E$ and $W_{1\perp}^E$ are the energy stored in the longitudinal and transverse components of electric field $\vec{E}$ in the sapphire disks with a sample between them; $\tan\delta_\parallel$, $\tan\delta_\perp$ are the components of the loss tangent tensor of sapphire in the direction of optic axis and in the plane perpendicular to this axis respectively.

Similarly to (6) and (7) the coefficients $K_{1\parallel}$ and $K_{1\perp}$ are calculated via integration of longitudinal and transverse components of vector $\vec{E}$ or by numerical differentiation of the resonant frequency dependences:

$$K_{1\parallel} = -2\frac{\varepsilon_{1\parallel}}{\omega}\left(\frac{\partial\omega}{\partial\varepsilon_{1\parallel}}\right), \qquad K_{1\perp} = -2\frac{\varepsilon_{1\perp}}{\omega}\left(\frac{\partial\omega}{\partial\varepsilon_{1\perp}}\right). \tag{9}$$

The relations $K_{1\parallel} >> K_{1\perp}$ for $HE_{n11}$ modes and $K_{1\parallel} << K_{1\perp}$ for $EH_{n11}$ modes are typically valid. The relations $K_{1\parallel}, K_{1\perp} > 0$ and $0 < K_{1\parallel} + K_{1\perp} < 1$ are satisfied.

*C. SURFACE RESISTANCE MEASUREMENTS UNDER DIELECTRIC*

To measure the surface resistance $R_S$ of the metallic foil on its interface with the dielectric material, the disk (1) is pressed against unclad surface of the sample (Fig. 1a and Fig. 7d below). In order to determine $R_S$ we will use values of the filling factor $K_{3E}$ for laminated sample and the unloaded quality factor $Q_{0DR}$ of the resonator as well as $\tan\delta$ of an unclad sample measured in accordance with the procedure described in the previous section. Similarly to (4) and taking into account Eq. (5) the unloaded quality factor $Q_{0\Sigma\sigma}$ of the resonator pressed onto the dielectric sample with copper laminated layer is defined as

$$\frac{1}{Q_{0\Sigma\sigma}} = \frac{P_{\Sigma\sigma}}{\omega W_\Sigma} = \frac{1}{Q_{0DR}} + K_{3E}\cdot\tan\delta + \frac{1}{Q_\sigma}, \qquad (10)$$

where $P_{\Sigma\sigma}$ is the total loss power; $Q_\sigma = \left(\frac{P_\sigma}{\omega W_\Sigma}\right)^{-1}$ is a partial $Q$ factor due to ohmic loss in the metallic foil; $P_\sigma$ is an ohmic loss power, which is equal to

$$P_\sigma = \frac{R_S}{2}\int_S |H_\tau|^2 dS \quad , \qquad (11)$$

where $H_\tau$ is the tangential component of the microwave magnetic field on the surface of the metal; $S$ is the surface area at the interface between the metallic foil and the dielectric layer.

Total energy $W_\Sigma$ stored in the resonator can be found by integrating magnetic field energy in partial resonator volumes

$$W_i^H = \frac{\mu_0\mu_i}{2}\int_{V_i}\left(|H_{i\perp}|^2 + |H_{iz}|^2\right)dV \ . \qquad (12)$$

Neglecting the contribution of the longitudinal component of the magnetic field in Eq. (12) we get from Eqs. (10), (11), (12)

$$R_S = \frac{\pi f_n \mu_0}{Q_\sigma}\cdot M\ , \qquad (13)$$

where

$$Q_\sigma^{-1} = Q_{0\Sigma\sigma}^{-1} - Q_{0DR}^{-1} - K_{3E}\cdot\tan\delta\,,$$

$$M = \frac{4\sum_i W_i^H}{\mu_0\int_S |H_{3\perp}|^2 dS} = \frac{2\sum_i \mu_i\int_{V_i}|H_{i\perp}|^2 dV}{\int_S |H_{3\perp}|^2 dS} = \frac{2\sum_i \mu_i\int_S |H_\perp(r,\varphi)|^2 dS\int_l Z_{iE}^2(z)dz}{\int_S |H_\perp(r,\varphi)|^2 dS\cdot Z_{3E}^2(0)} \quad ,$$

where functions under integrals are further defined in Appendix A. For non-magnetic materials ($\mu_i = 1$) the geometric factor $M$ in Eq. (13) can be written as:

$$M = \left( \frac{\varepsilon_{1\perp} A_1}{\varepsilon_3} \right)^2 I_1 + \left( \frac{\varepsilon_2 A_2}{\varepsilon_3} \right)^2 I_2 + I_3 + \left( \frac{\varepsilon_2 A_4}{\varepsilon_3} \right)^2 I_4 \ .$$

The expressions for $A_i$ , $I_i$ are given in Appendix A.

## III. EXPERIMENT

### *A. MEASUREMENT CELL AND SETUP DESCRIPTION*

A simplified schematic of the measurement cell is shown in Fig. 3. A sample is placed between polished sapphire disks with diameter 12.51 mm and height 2.54 mm, which are arranged inside a thick-wall aluminium shield with inner diameter of 25 mm. The diameter of the shield was chosen to exclude any influence of the metal wall on either the resonant frequencies or the quality factors of the sapphire disks. The aluminium shield is placed inside a thermal isolation chamber. The lower sapphire disk is attached to the post guide and clamped to the sample through the spring.

We took special care to prepare "nearly ideal" dielectric resonators. Sapphire disks were cut from the same piece of a carefully oriented sapphire single crystal of very high chemical purity. The dimensions of both disks were identical to an accuracy of within 1 μm. The c-axis of both the disks was perpendicular to their faces. The faces of each disk were parallel with the accuracy better than 1 μm across the disks diameter. Surface roughness reduced to 2 nm after polishing. The deviation from flatness of each surface was less than 0.5 μm across each disk's diameter.  Both disks were oriented crystallographically using X-ray scattering.

As a result the problem of the "residual air-gap" was significantly reduced even when the two disks were brought into contact without a "soft" dielectric film between them. Moreover, because these two disks in close mechanical contact constitute a near-perfect monolithic crystal, no measurable splitting of the resonance curves has been detected.  It is well known that splitting of the resonance curve could result from the asymmetry of the dielectric disks in a split dielectric resonator.  Such parasitic splitting could introduce a source of additional error in relation to measurements of very thin dielectric films. The sophisticated technology used for the preparation of the "near-ideal" disks of the sapphire split resonator helps to significantly reduce such problems.

Special care was also taken in the mechanical design of the test rig in order to allow measurements at variable temperatures. Because of the strong temperature dependence of sapphire permittivity [17] parasitic temperature gradients otherwise might influence the interpretation of the temperature dependence measurements of thin film dielectric properties.

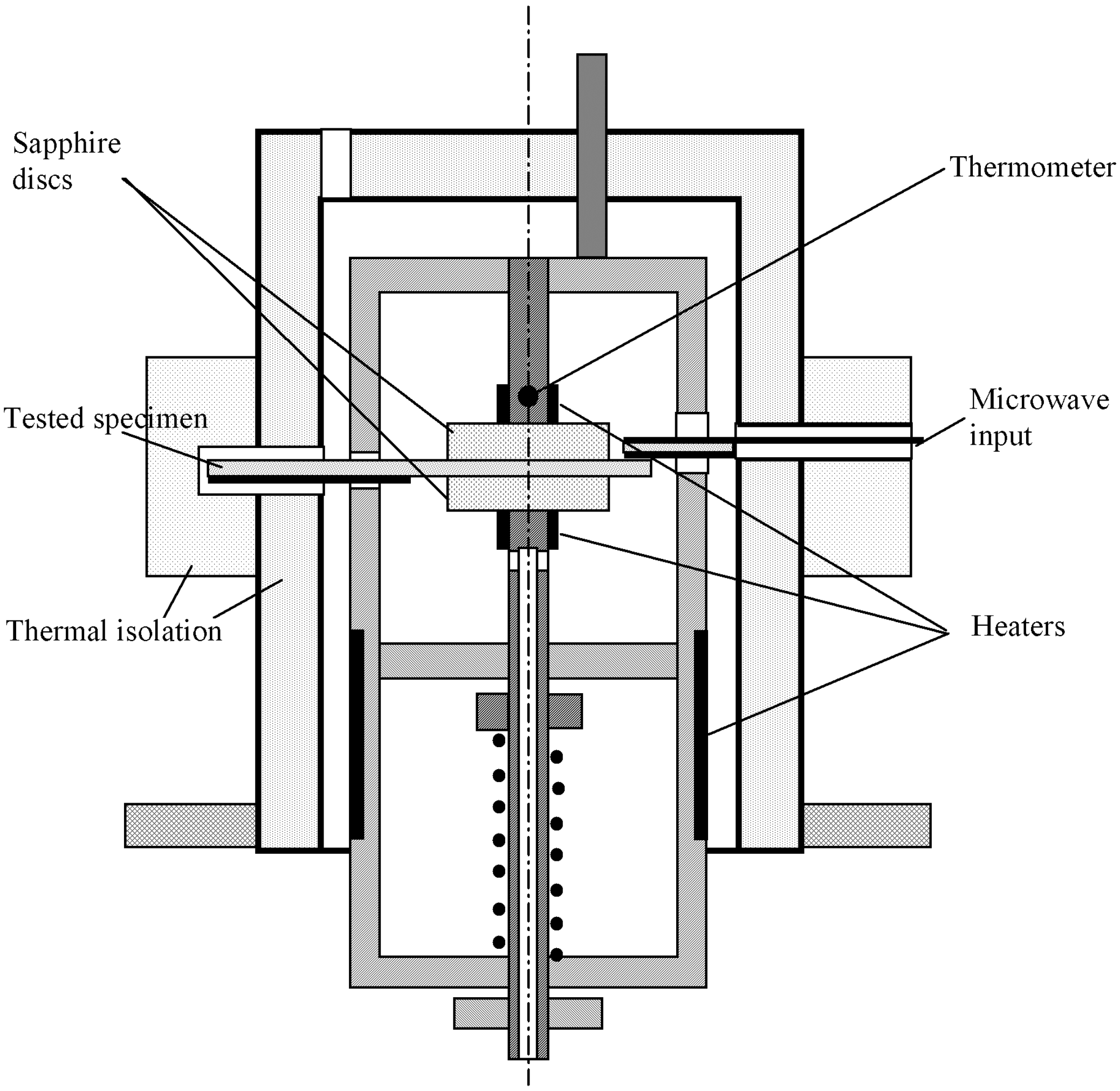

Fig. 3. Measurement cell diagram

To simplify the process of changing dielectric samples the post guide is designed to be axially moveable and have no radial free play. The aluminium shield with sapphire disks and test sample can be moved towards and away from the microwave excitation microstrip line by a stepper-motor (not shown) in order to tune the resonator coupling to the excitation line. Semi-rigid coaxial cables connect the microstrip to standard 2.9 mm connectors outside the cell thermal isolation. The measuring cell is placed in a stainless steel vacuum cryostat with a temperature control system. For low temperature measurements, liquid nitrogen is evaporated from the cryostat and its vapour flows around the resonator and the aluminium shield.

The block-diagram of the measurement setup is shown in Fig. 4. A photo of the complete test rig setup is shown in Fig. 5. It includes: a signal generator Rohde&Schwarz SMR40 (№1); Agilent 82357A USB/GPIB adapter (№2) for connecting the generator to the computer; microwave cables (№3); the main part of the measurement cell (№4); electric power supply (№5) for temperature control (in the range from -50 up to +70 $^{0}$C) of the resonator and some of the samples used for measurements (№6).

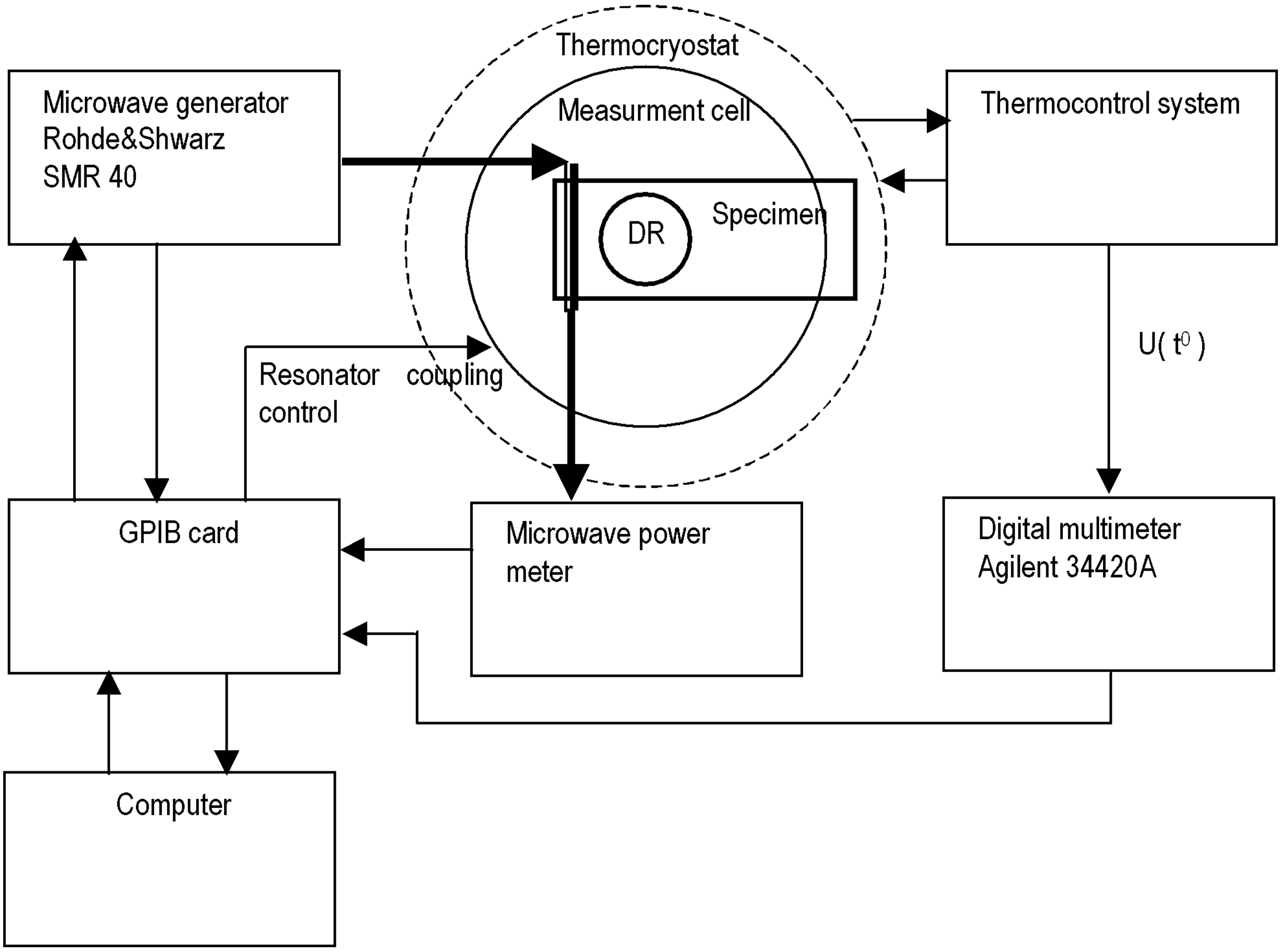

Fig. 4. The block-diagram of measurement setup

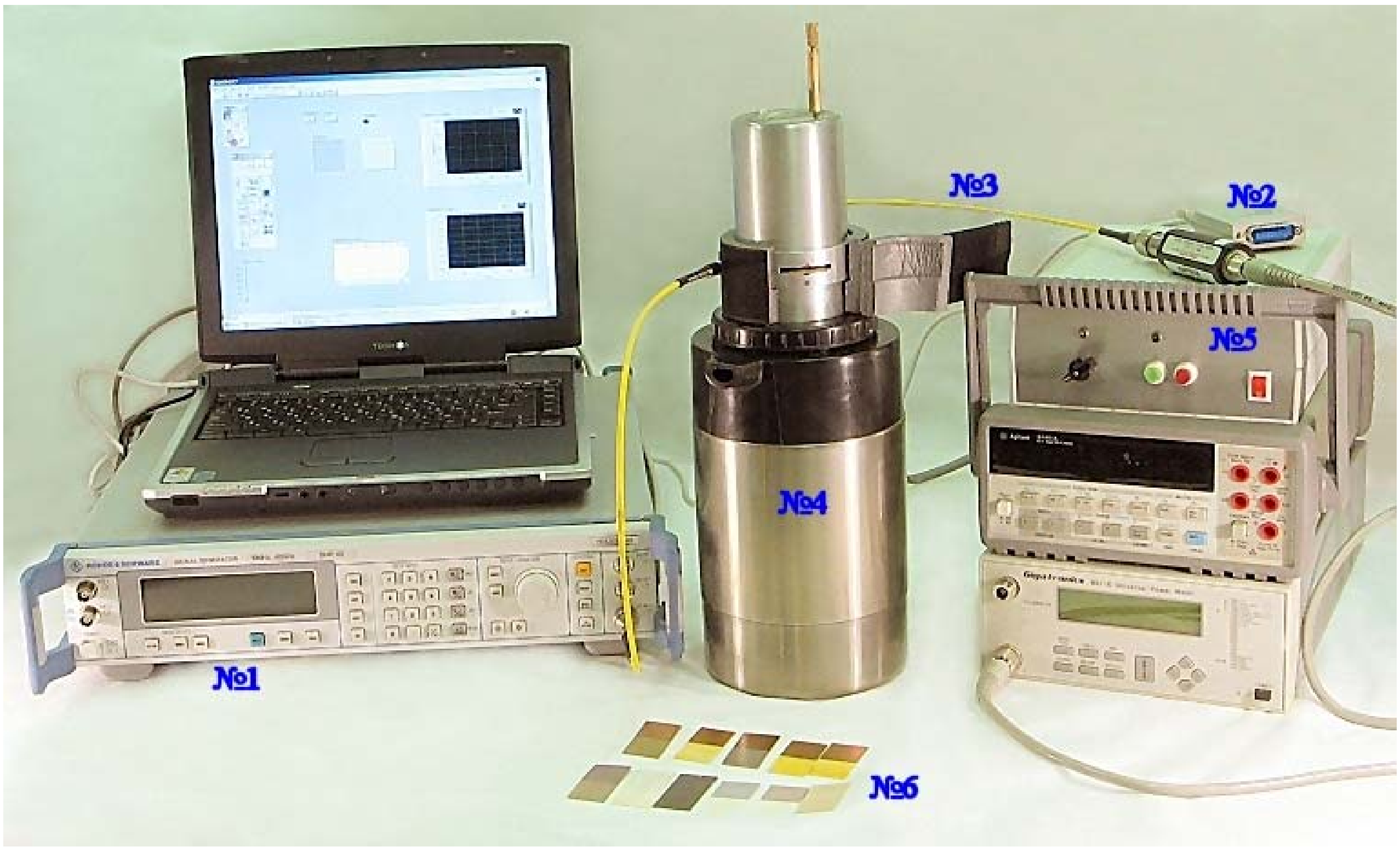

Fig. 5 Measurement setup photograph

### *B. EXPERIMENTAL PROCEDURE AND RESULTS*

The procedure for taking measurements of dielectric constant, loss tangent and surface resistance of one-side laminated dielectric samples is described below.

First the resonant spectrum (resonator output microwave power *P* vs. frequency *f*) of the upper dielectric resonator is measured. For this measurement the lower resonator is moved away by a maximum distance of 3mm from the upper resonator and does not influence the measured quantities. Thereupon we determine the resonant frequencies $f_n$, loaded quality factors $Q_n$, and coupling coefficients $\beta_n$ of $HE_{n11}$ modes ($8 \le n \le 12$) of the upper resonator in the range 30 ≤ $f_n$ ≤ 40 GHz. An example of the frequency spectrum measured at room temperature $T$ = 22 $^0$C for a single sapphire disk is shown in Fig. 6. The caption to Fig. 6 shows *P(f)* experimental curve corresponding to the $HE_{811}$ mode (circles) and the results of the Lorentzian curve fit (solid line). The unloaded quality factor $Q_8$ = 35790 for this mode was obtained by the formula $Q_8 = Q_L(1+\beta_8)$, where $Q_L$ is the measured (loaded) quality factor. The unloaded quality factors of the $HE_{n11}$ modes with $n$ =9, 10, 11, 12 are equal to 40850, 45360, 44970, and 37080 respectively. The maximum quality factor corresponds to $HE_{10,1,1}$ mode. The further increase of azimuth index *n* results in a drop in $Q_n$ due to the increase of the sapphire loss tangent. The latter is roughly proportional to the frequency.

The measured values of resonant frequencies and quality factors for $HE_{n11}$ modes of the single resonator at different temperatures ñ50 ≤ $T$ ≤ 70 $^0$C are saved into computer memory as calibration constants for further calculations.

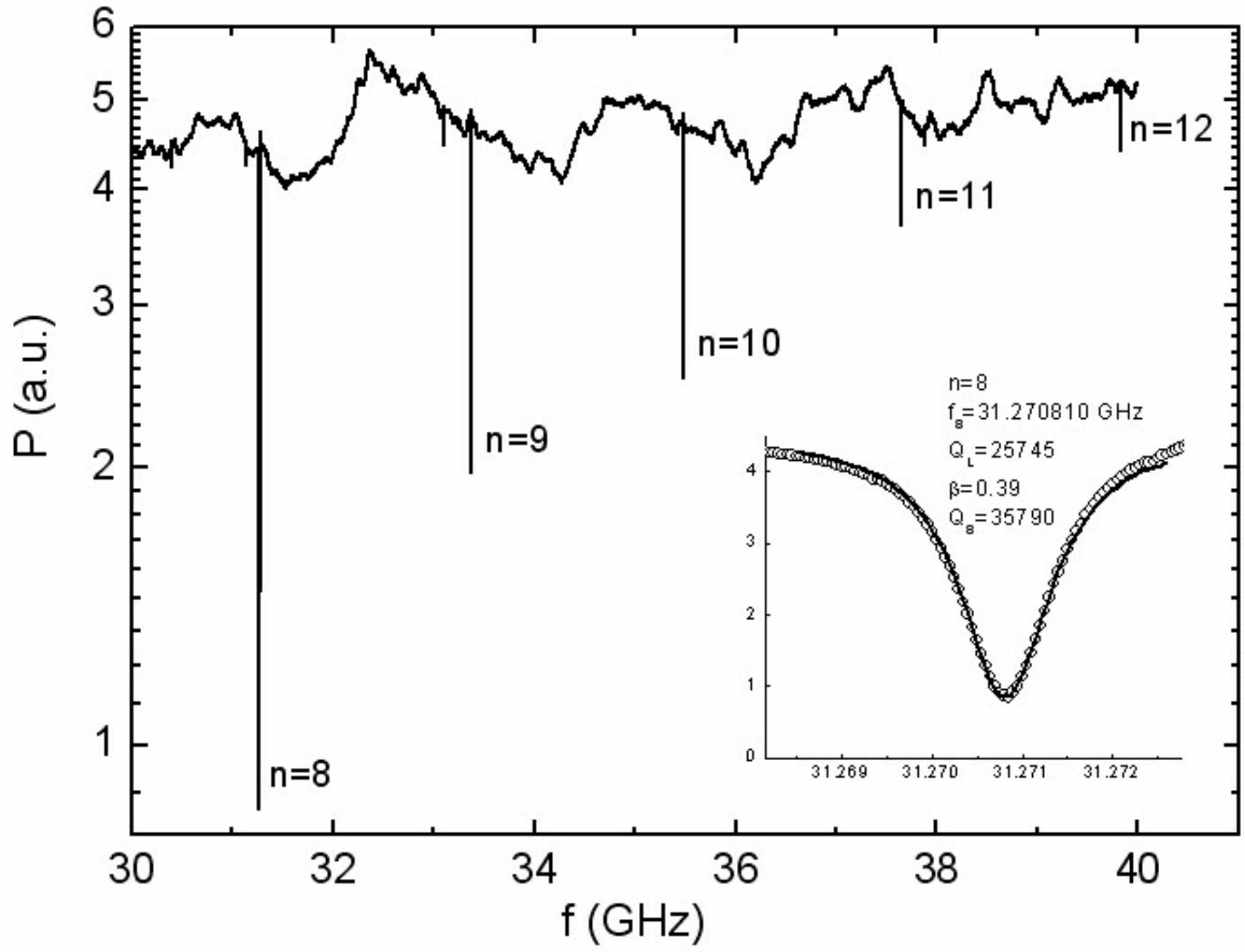

Fig. 6. Resonant spectrum of the single sapphire disk

We proceed with similar measurements with both sapphire disks pressed together and determine the values of resonant frequencies and quality factors for $HE_{n11}$ modes of this doubled resonator at the same temperatures ñ50 ≤ $T$ ≤ 70 $^0$C. The results obtained from spectra are also stored into the computer memory for future calculations of the dielectric constant, loss tangent and surface resistance of laminated dielectric samples.

The measured frequencies $f_n^2$ of the double resonator are significantly lower than corresponding frequencies $f_n^1$ of the single one. The difference ($f_n^1$ - $f_n^2$) decreases when the azimuth number $n$ increases. For example, it is equal to 4406 MHz for $n$ = 9 and 3637 MHz for $n$ = 12. This approximately corresponds to the theoretical calculations for the double resonator. Results of theoretical calculations using Eqs. (1) and (2) are shown in Table 1 along with the measured resonant frequencies. In these calculations we used sapphire permittivities of $\varepsilon_{1\|} = 11.577$ and $\varepsilon_{1\perp} = 9.388$ [17]. The relative discrepancy $\delta f = (f_{calc} - f_n)/f_n$ between the calculated $f_{calc}$ and measured $f_n$ frequency values does not exceed 1.3 % for the single resonator and 0.5 % for the double resonator. This discrepancy does not affect further results as the experimentally determined shift of the frequency for resonator with and without sample is used as the input for calculations.

Table 1. Theoretical and experimental resonant frequencies of $HE_{n11}$ modes

| | Single resonator | | | Doubled resonator | | |
|---|---|---|---|---|---|---|
| $n$ | $f_{calc}$, GHz | $f_n^1$, GHz | $\delta f \cdot 10^2$ | $f_{calc}$, GHz | $f_n^2$, GHz | $\delta f \cdot 10^2$ |
| 9 | 32.9767 | 33.3873 | -1.23 | 28.8439 | 28.9808 | -0.47 |
| 10 | 35.1495 | 35.4969 | -0.98 | 31.2617 | 31.3881 | -0.40 |
| 11 | 37.3448 | 37.6594 | -0.84 | 33.6768 | 33.7958 | -0.35 |
| 12 | 39.5590 | 39.8415 | -0.71 | 36.0889 | 36.2041 | -0.32 |

The procedure for the measurement of the dielectric constant and loss tangent of an unclad thin dielectric sample is described in parts A and B of Section II. Dielectric sheet samples used for measurements have planar dimensions 25x50 mm$^2$, thickness up to 1mm and one-side copper laminated square surface of 25x25 mm$^2$. The sample is held with force between the upper and lower resonators (Fig. 7a) providing the sapphire disks are in the centre of the square 25x25 mm$^2$ surface.

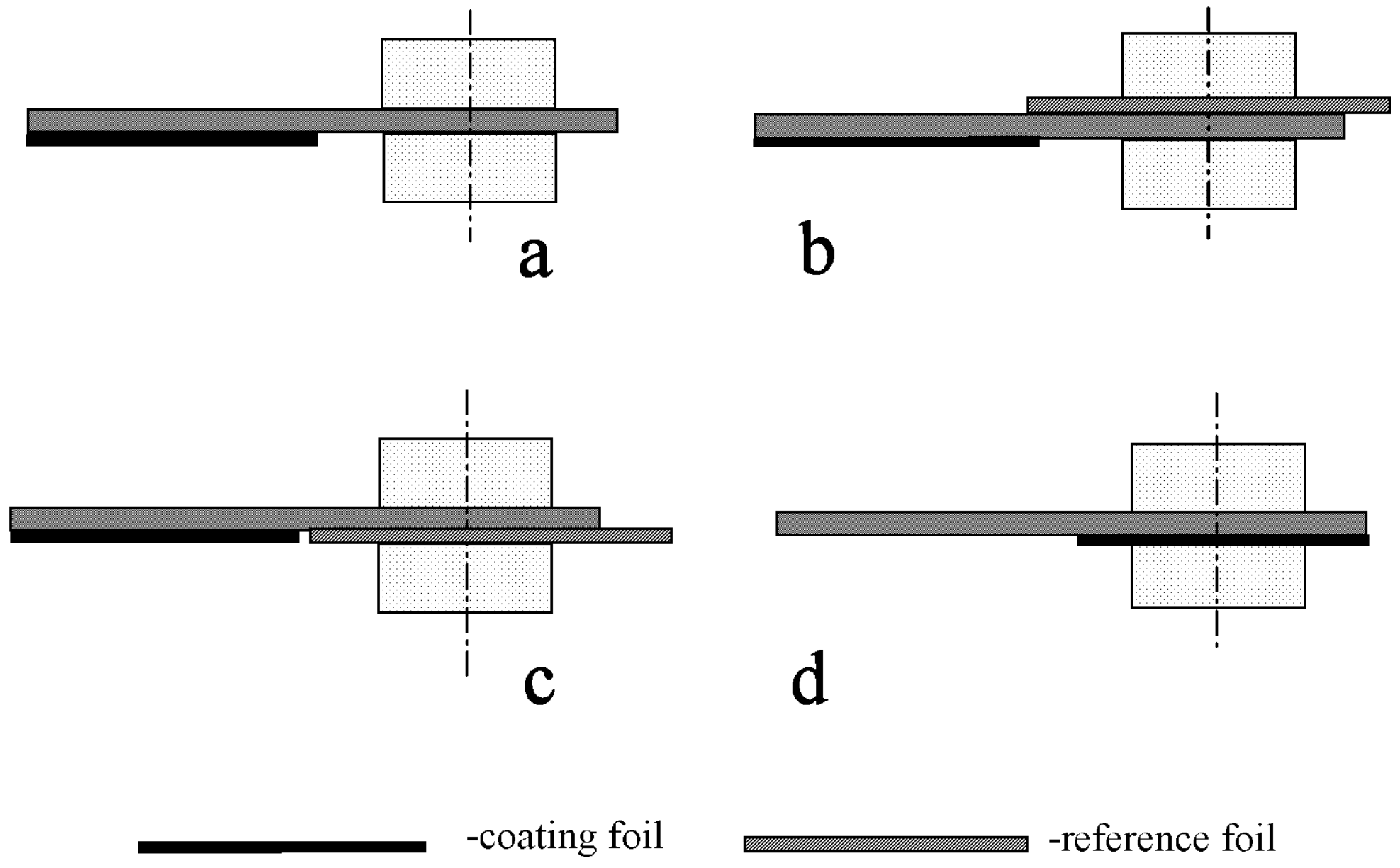

Fig. 7. Schemes for measurements of: a) $\varepsilon$, $\tan\delta$; b) $R_{Sref}$; c) $Q_{ref}$; d) $Q_\sigma$, $R_S$.

When the sample is clamped inside the split dielectric resonator, the measured resonant frequencies $f_{n\varepsilon}$ are shifted down compared to the frequencies $f_n$ of the single resonator. The problem of identification of the $HE_{n,1,1}$ mode arises. Fortunately, however, first the coupling of the $HE_{n,1,1}$ modes are the highest, and secondly the frequency difference $\Delta f = f_{n,\varepsilon} - f_{n-1,\varepsilon}$ between the nearest $HE_{n,1,1}$ and $HE_{n-1,1,1}$ modes is almost independent for $n \geq 9$ as shown in Fig. 8.

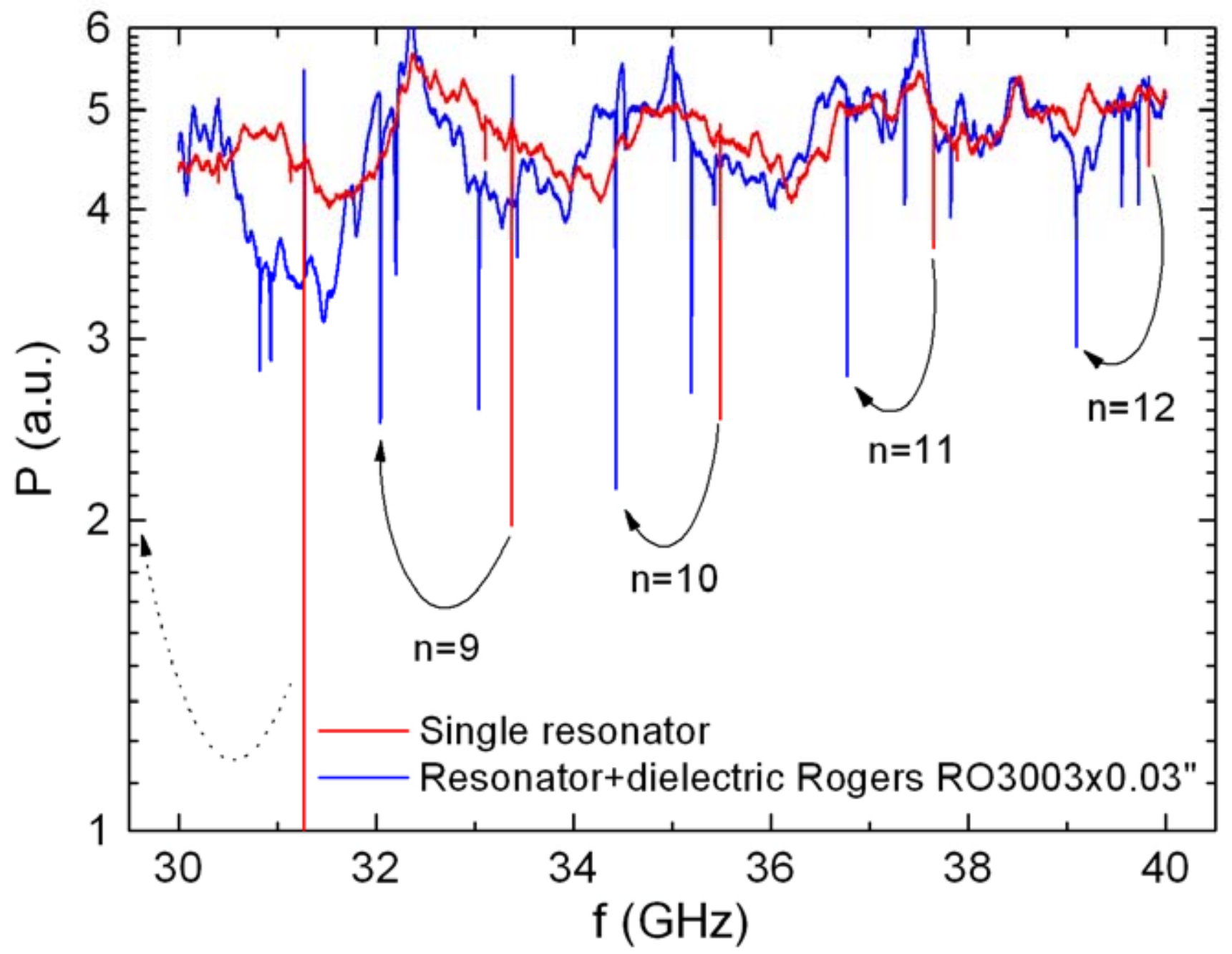

Fig. 8. Resonant spectra of a single resonator and dielectric sample RO3003x0.03".

Actually, using an approximate (with uncertainty of 20%) value of $\varepsilon$, which can be found from low-frequency capacity measurements, from Eqs. (1) and (2) we calculate the approximate frequency $f_{n,\varepsilon}^{\text{appr}}$ for a particular mode with an index *n* and perform the frequency sweep in the range $f_{n,\varepsilon}^{\text{appr}} \pm 200\text{MHz}$ looking for the most deep resonance. This procedure is successively repeated for higher $HE_{n,1,1}$ modes. Immediately after the measurement at a particular mode is finished, the resonance curve is fitted in order to obtain resonance frequency, loaded quality factor, coupling coefficient, and unloaded quality factor. If necessary, the coupling between the resonator and the microwave line is changed to obtain the optimal sensitivity. In such case measurements with a new coupling are repeated. Using Eqs. (1) - (9) the values of permittivity and loss tangent are calculated. If the mode identification was correct, the results of calculations for all the modes give very close values for both the permittivity and the loss tangent. Finally, the mean values of the permittivity and the loss tangent are calculated using results obtained for all modes. The results are weighted according to the error of the resonance curve fitting. Examples of such results obtained for a set of samples at room temperature are shown in Table 2 (Appendix B). Below the mean values of permittivity and loss tangent coefficients $C_\varepsilon = (\partial\varepsilon/(\varepsilon \cdot \partial t)) \cdot 1000$ and $C_\delta = (\partial(\tan\delta)/(\tan\delta \cdot \partial t)) \cdot 1000$ are shown. These values are introduced as correction coefficients describing the influence of the absolute uncertainty $\Delta t$ (in µm) in measurements of the thickness of the sample. The error for permittivity can then be found by the formula $\Delta\varepsilon/\varepsilon = C_\varepsilon \cdot \Delta t/1000$. Similarly the error in the loss tangent value is given by $\Delta(\tan\delta)/\tan\delta = C_\delta \cdot \Delta t/1000$.

The test rig described above allows measurements to be performed at different temperatures. Such measurements are similar to those at room temperature. The only difference is that preliminary calibration of the resonance frequencies and quality factors of the single and double resonator are performed in the range of temperatures. The temperature is stabilised at exactly the same values for measurements with and without the sample, which helps to compensate almost completely for the temperature dependence of sapphire dielectric properties. An example of temperature dependences obtained by this scheme for two samples - NY9220x0.01" and Tly5a0200 - is presented in Fig. 9 and Fig. 10 for the permittivity and tangent loss respectively.

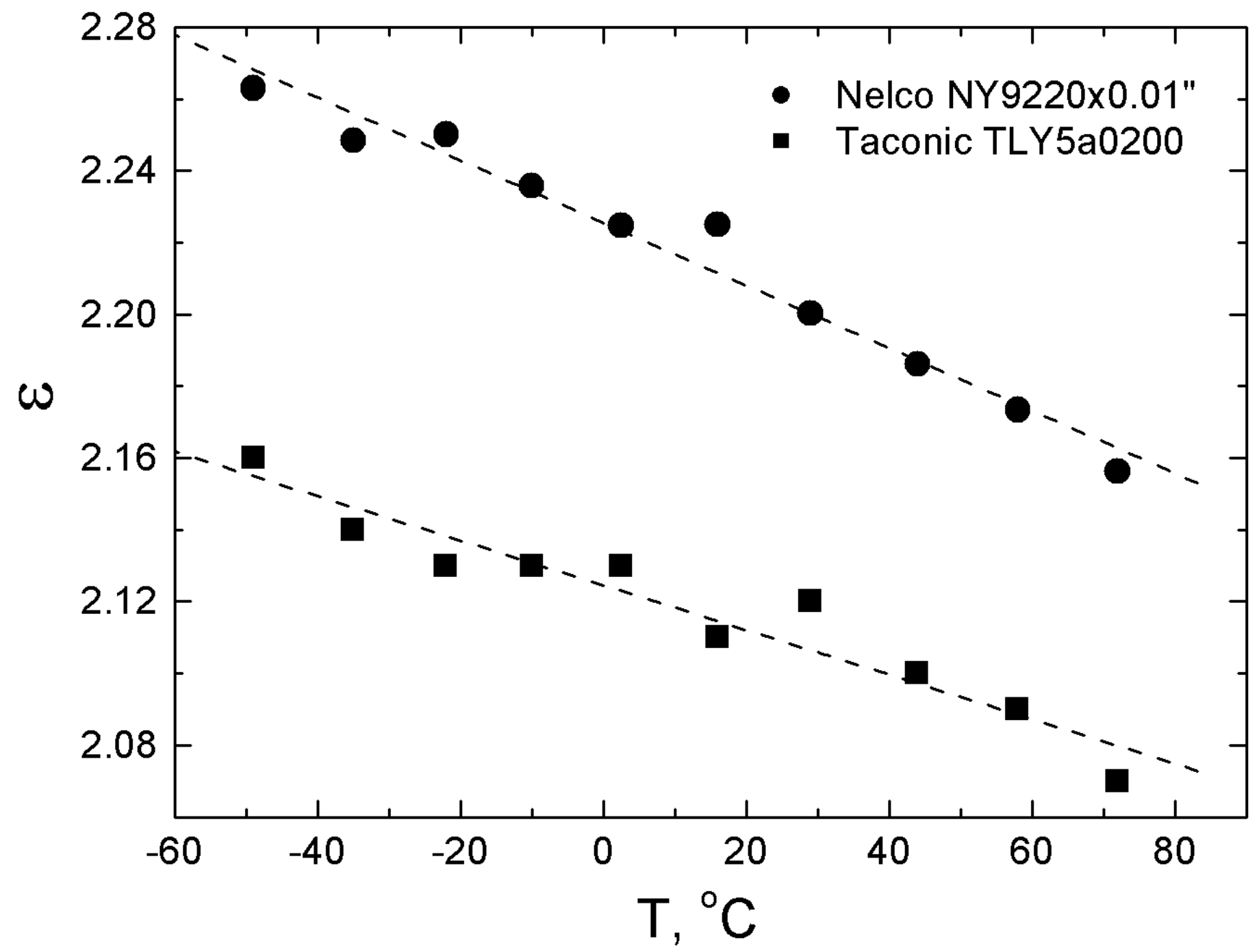

Fig. 9. Temperature dependences of permittivity in NY9220x0.01" (Nelco) and Tly5a0200 (Taconic) samples.

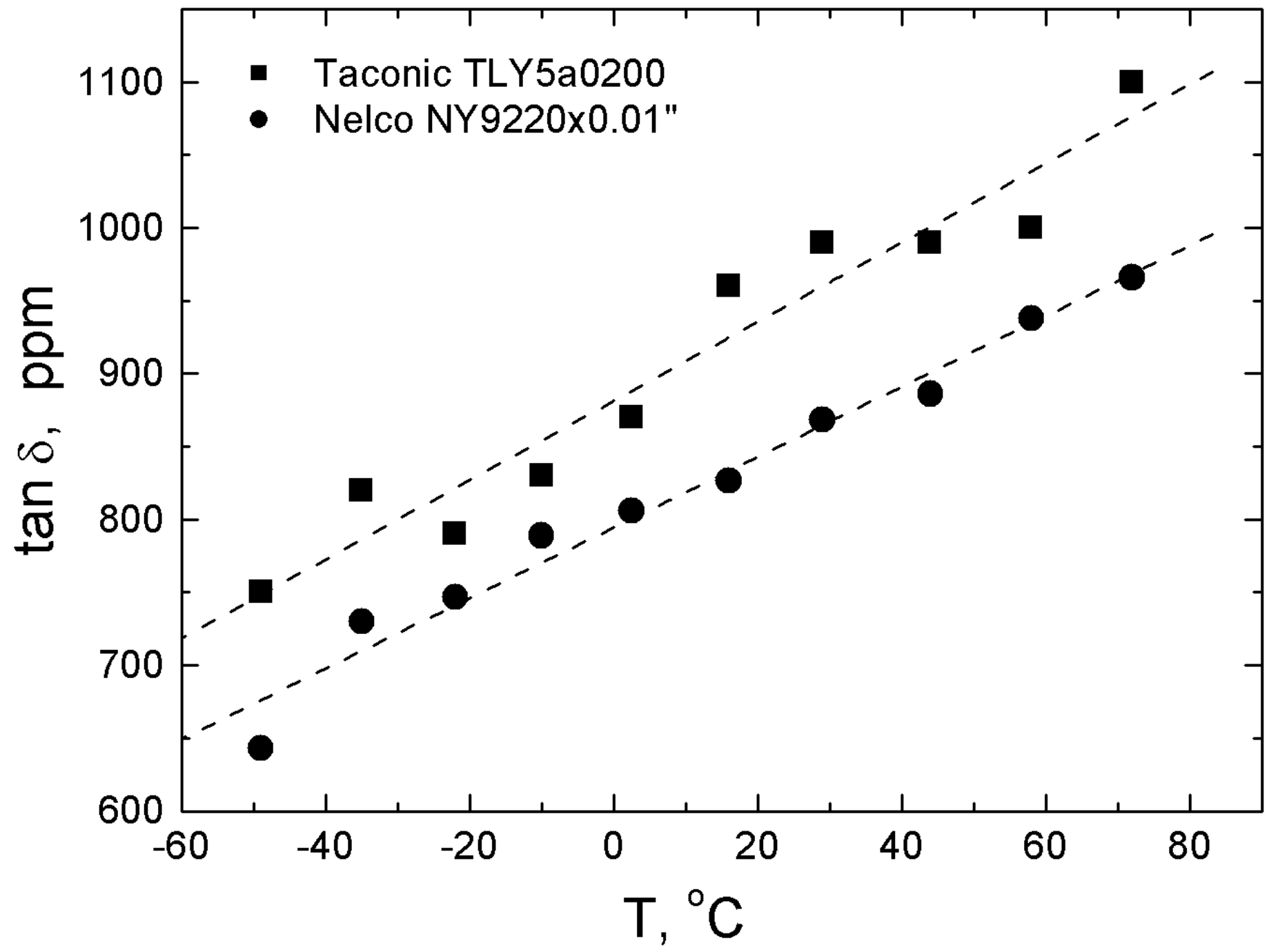

Fig. 10. Temperature dependences of loss tangent in NY9220x0.01" (Nelco) and Tly5a0200 (Taconic) samples.

The method for making surface resistance $R_S$ measurements of the laminated dielectric samples is illustrated in Figs. 7b-7d. Two approaches are possible here: (i) direct measurements and (ii) measurements using calibrated reference copper foil. Let us consider them separately.

(i) The direct method of the surface resistance measurements is based on the calculation of $R_S$ using Eq. (13). In this case the sample is placed into the resonator as shown in Fig. 7d and the quality factor $Q_{0\Sigma\sigma}$ of the upper resonator with laminated dielectric sample is measured. Using the previously determined quality factor $Q_{0DR}$ of the resonator without a sample and the sample loss tangent $\tan\delta$ (obtained by measuring the non-clad part of the sample as described above) the value of $Q_\sigma$ is obtained:

$$Q_\sigma = \left(Q_{0\Sigma\sigma}^{-1} - Q_{0DR}^{-1} - K_{3E} \cdot \tan\delta\right)^{-1} \qquad (14)$$

where filling factor $K_{3E}$ is calculated for laminated sample. Factor $Q_\sigma$ has appeared before as a denominator in the right side of Eq. (13). It characterises an ohmic loss in the metal lamination at the interface with dielectric material. The geometric factor *M* is calculated from Eq. (13).

Fig. 11 illustrates the frequency shifts and the quality factor variations for measurements at the fixed mode with azimuth number ten for a particular sample. The measurements on the laminated part of the sample were used for direct measurements of $R_{S.}$

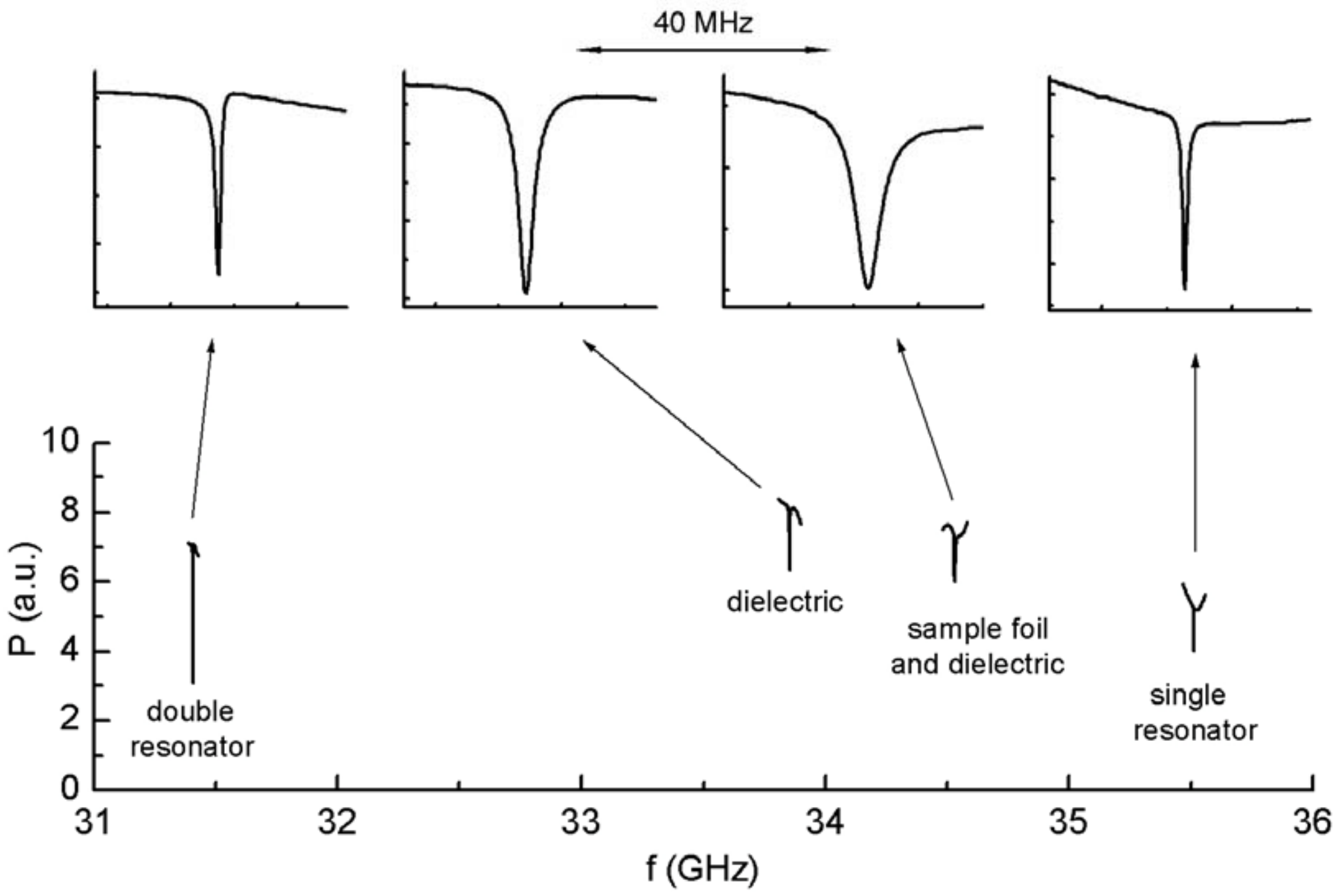

Fig. 11. Evolution of resonance curves of $HE_{10,1,1}$ mode in the sample RO3003x0.03" (Rogers) for direct measurement of surface resistance.

The accuracy of the direct surface resistance measurements strongly depends on the thickness *t* and the dielectric losses in the substrate. To obtain reliable results by this method ohmic losses in the laminated metal must be comparable to dielectric losses. In case of copper foil the applicability criteria for direct measurements can be written as $t < 0.1 \cdot (\varepsilon / \tan\delta)$, where the thickness *t* of the substrate is expressed in μm. In Table 2 (Appendix B) the $R_S$ data of RO3003x0.01", GIG M/L-2, NH9348ST0203CHCH, Rogers 3003, Rogers 5880, Sheldahl G2200x2mil and Sheldahl Comclad XFx10mil samples were obtained by direct measurements.

(ii) The second method to measure the surface resistance of laminated dielectric samples involves a few extra steps, which are shown in Figs. 7b-7d. These steps are described in more detail below:

1. The smooth copper metal foil is chosen as a reference. The surface resistance of the foil $R_{Sref}(f)$ (if unknown) is determined by measuring the quality factor of the upper resonator pressed against this foil (Fig. 7b). Using measurements at the resonant frequencies of several $HE_{n,1,1}$ modes the surface resistance can be calculated as follows:

$$R_{Sref} = \frac{\pi f_n \mu_0}{Q_{\sigma ref}} \cdot M_{ref} . \qquad (15)$$

where $Q_{\sigma ref}^{-1} = Q_{0\Sigma\sigma}^{-1} - Q_{0DR}^{-1}$, $M_{ref} = \left( \frac{\varepsilon_{1\perp} A_1}{\varepsilon_2} \right)^2 I_1 + I_2 + A_4^2 I_4$. The expression for $M_{ref}$ follows from Eq. (13) for the resonator located on a metallic plane without dielectric layer ($t = 0$) but with an effective air-gap *d*. In turn, effective air-gap can be found from Eqs. (1) and (2) and the condition of equality of double resonator frequency to the frequency of the upper resonator on the metal surface. To obtain $A_i$ and $I_i$ values from Eqs. (A6) of the Appendix A one should use $t = 0$ and obtained air-gap value *d*.

2. The reference foil is placed underneath the uncoated region of the dielectric sample, and they are held together between the disks of the split resonator (Fig. 7c). The quality factors $Q_{0\Sigma\sigma ref}$ of this sandwich structure are measured for different resonance $HE_{n,1,1}$ modes. The values $Q_{ref}$ at resonant frequencies are founded by the formula:

$$Q_{ref} = \left( Q_{0\Sigma ref}^{-1} - Q_{0DR}^{-1} - K_{3E} \cdot \tan\delta \right)^{-1} . \qquad (16)$$

In contrast to $Q_{\sigma ref}$ determined during the first step, the quality factors $Q_{ref}$ in Eq. (16) determine the losses in the reference foil taking into account the electromagnetic field distribution in the structure of Fig. 7c.

3. The same distribution of the field occurs in the geometry shown in Fig. 7d. The quality factor $Q_\sigma$ related to the ohmic loss at the interface between the metal foil and the dielectric material is determined in accordance with Eq. (14). The surface resistance $R_S$ of the metal foil at the resonant frequencies of $HE_{n,1,1}$ modes is found as

$$R_S = R_{Sref} \frac{Q_{ref}}{Q_\sigma}, \tag{17}$$

where the value $R_{Sref}$ of the reference foil measured at the first step is linearly approximated to the appropriate frequency of the third measurement step.

The advantage of this method (ii) in comparison with the direct one (i) is that the surface resistance $R_S$ does not depend on the calculation of the geometric factor $M$ in Eq. (13) and, hence, the accuracy of method (ii) is higher, especially for thicker samples.

## VI. Conclusion

In this paper we present a novel technique for the measurement of the dielectric constant and loss tangent of dielectric substrates with reasonable accuracy for substrate thickness ranging from 10 to 1000 μm. For the first time a resonance technique with the electric field of electromagnetic radiation orthogonal to the surface of the substrate has been demonstrated. The high sensitivity for very thin samples is made possible by the high unloaded quality factor of the "whispering gallery" resonator and substantial filling factor value even at very small substrate thickness. The main inaccuracy for microwave measurements on thin dielectric films with our method is determined by the uncertainty in the sample's thickness. There is no fundamental restriction on the maximum thickness of substrate while its dielectric permittivity is lower than the one of sapphire. When the dielectric thickness increases, the measurement structure shown in Fig. 1 gradually turns to single DR on the dielectric half-space. In such cases a single DR resonance spectrum represented in Fig. 4 will just shift to lower frequencies. The contribution of the second substrate face will finally become negligible due to the exponential decay of the electromagnetic field in the substrate.

Experimental results do not show any influence of the "residual air-gap" problem, which is explained by the optical-quality sapphire polishing, elasticity or/and flatness of most of the samples, as well as by pressure applied between sapphire disks and the substrate. Nevertheless it was found that the measurement set-up is sensitive to contamination of the surface of the test samples. We have therefore to limit the presence of dust or fibres between the surfaces of the substrate and the sapphire disks. The value of the frequency of the resonant modes of the double sapphire resonator without a sample was used as a control parameter related to the sapphire face contamination. When necessary the faces of the split sapphire resonator were cleaned with special tissue. These peculiarities are inherently caused by the orthogonality of the electric field to the surface of the sample.

The method described also provides with reasonable accuracy measurements of the surface resistance of metal films. The presence of copper film in the resonator reduces the quality factor by an order of magnitude. The accuracy of surface resistance measurements at the interface between a metallic film and a dielectric layer is strongly influenced by the substrate thickness, dielectric constant and loss tangent. In the case of $\varepsilon \geq 2$ and $\tan\delta \sim 10^{-4}$ the 15-20% accuracy of $R_S$ measurements was shown experimentally for dielectric substrate thickness $t \leq 0.2 \div 0.5$ mm. Such materials are widely used in 30-40 GHz frequency range.

## ACKNOWLEDGMENT

We would like to acknowledge the support of the National Measurement System Directorate of the UK Department of Trade and Industry for our development of this novel microwave measurement technique. We would like especially to thank Rhode&Schwartz for providing a long-term loan of a microwave signal generator. We also want to use this chance to express our gratitude to the many companies which have provided samples for our validation measurements. This list includes, but is not limited to, Rogers, Labtech, Spemco, Isola, Sheldahl, Nelco, Bookham Technology, and Celestica. The authors particularly wish to thank S.V. Ryzhkov for help with sample preparation, V.N. Kurlov for assistance with sapphire disk polishing, and G.E. Tsydynzhapov for assistance with computer automation of the test procedures.

## APPENDIX A

We represent an electromagnetic field inside the resonator (Fig. 1a) within the boundaries $t+d \leq z \leq t+d+L$ as a hybrid standing $HE$ or $EH$ wave along the $z$-axis. The transverse field distribution in the gap (2), dielectric layer (3) and top space (4) in Fig. 1b is assumed the same as in the disk (1). A set of Maxwell equations result in a corresponding set of wave equations for longitudinal field components inside the dielectric cylinder $r \leq a$ (the space marked as (+) in Fig. 1) [14]

$$\nabla^2\Psi_{iE}^{+} + k_i^2\left(\varepsilon_{1\parallel}/\varepsilon_{1\perp}\right)\Psi_{iE}^{+} - \left(1-\frac{\varepsilon_{1\parallel}}{\varepsilon_{1\perp}}\right)\cdot\frac{\partial^2\Psi_{iE}^{+}}{\partial z^2} = 0, \quad \text{(A1)}$$

$$\nabla^2\Psi_{iH}^{+} + k_i^2\Psi_{iH}^{+} = 0,$$

where $\Psi_{iE}^{+} = E_{iE}^{+}(r,\varphi,z)$, $\Psi_{iH}^{+} = H_{iH}^{+}(r,\varphi,z)$ and $i = 1, 2, 3, 4$ are the numbers for dielectric layers along the $z$-axis. In the space outside the dielectric cylinder, $r > a$, marked as (-) in Fig. 1, the longitudinal field components should satisfy the following condition:

$$\nabla^2\Psi_{iE,H}^{-} + k_2^2\Psi_{iE,H}^{-} = 0, \quad \text{(A2)}$$

where $\Psi_{iE}^{-} = E_{iE}^{-}(r,\varphi,z)$, $\Psi_{iH}^{-} = H_{iH}^{-}(r,\varphi,z)$, and $i = 1, 2, 3, 4$.

We look for the solutions of Eqs. (1), (2) in the form of:

$$E_{iz}^{\pm} = R_E^{\pm}(r)\cdot\Phi_E(\varphi)\cdot Z_{iE}(z)\,,$$

$$H_{iz}^{\pm} = jR_H^{\pm}(r)\Phi_H(\varphi)\cdot Z_{iH}(z)\,, \qquad i=1,2,3,4 \qquad \text{(A3)}$$

where $R_E^+(r) = A^+ J_n(\alpha\chi_1 r)$, $R_E^-(r) = A^+\cdot\dfrac{J_n(\alpha\chi_1 a)}{H_n^{(2)}(\chi_2 a)}\cdot H_n^{(2)}(\chi_2 r)$, $\Phi_E(\varphi) = \begin{pmatrix}\cos n\varphi\\ \sin n\varphi\end{pmatrix}$,

$$R_H^+(r) = B^+ J_n(\chi_1 r)\,,\quad R_H^-(r) = B^+\cdot\frac{J_n(\chi_1 a)}{H_n^{(2)}(\chi_2 a)}\cdot H_n^{(2)}(\chi_2 r)\;,\; \Phi_H(\varphi) = \begin{pmatrix}-\sin n\varphi\\ \cos n\varphi\end{pmatrix},$$

$A^+$, $B^+$ are the constants, $\chi_1$ and $\chi_2$ are the inner ($0\le r\le a$) and the outer ($a\le r<\infty$) transverse wave numbers respectively, $\alpha=\sqrt{\varepsilon_{1\parallel}/\varepsilon_{1\perp}}$. As $Z_{iE,H}(z)$ functions we choose the ones

$$Z_{iE}(z) = \begin{cases} A_1\cos(hz+\Theta_1) & (t+d\le z\le t+d+L)\\ A_i\cosh(h_{iE}z+\Theta_i),\ i=2,3; & (0\le z\le t;\ t\le z\le t+d)\\ A_4\exp(-h_{4E}z) & (t+d+L\le z<\infty)\end{cases}$$

$$Z_{1H}(z) = -A_1\sin(hz+\Theta_1)\,, \qquad (t+d\le z\le t+d+L)\,, \qquad \text{(A4)}$$

Neglecting the components $H_{iz}$ and their contribution in transverse components of $HE_{n11}$ modes within the gap (2) and the layers (3), (4):

$$Z_{iH}(z) = 0\;, \qquad i=2,3,4;\ (0\le z\le t;\ \ t\le z<t+d;\ \ t+d+L<z<\infty).$$

Here $A_i$ are the constants and $h_{iE}$ are the longitudinal wave numbers ($h_{4E}=h_{2E}$).

The transverse components of the resonator field are connected with the longitudinal ones by means of well-known relations [18]. The transverse mode field components will be expressed as:

$$E_{1r}^{\pm} = \frac{-A_1}{(\chi^{\pm})^2}\left[h\cdot\frac{\partial}{\partial r}R_E^{\pm}(r)\cdot\Phi_E(\varphi)+\frac{\omega\mu_0\mu_1}{r}R_H^{\pm}(r)\cdot\frac{\partial}{\partial\varphi}\Phi_H(\varphi)\right]\cdot\sin(hz+\Theta_1)\;,$$

$$E_{1\varphi}^{\pm} = \frac{-A_1}{(\chi^{\pm})^2}\left[\frac{h}{r}\cdot R_E^{\pm}(r)\cdot\frac{\partial}{\partial\varphi}\Phi_E(\varphi)-\omega\mu_0\mu_1\cdot\frac{\partial}{\partial r}R_H^{\pm}(r)\cdot\Phi_H(\varphi)\right]\cdot\sin(hz+\Theta_1)\;,$$

$$E_{ir}^{\pm} = \frac{-A_i}{(\chi^{\pm})^2}h_{iE}\cdot\frac{\partial}{\partial r}R_E^{\pm}(r)\Phi_E(\varphi)\sinh(h_{iE}z+\Theta_i)\;,$$

$$E_{i\varphi}^{\pm} = \frac{-A_i}{(\chi^{\pm})^2}\frac{h_{iE}}{r}\cdot R_E^{\pm}(r)\frac{\partial}{\partial\varphi}\Phi_E(\varphi)\sinh(h_{iE}z+\Theta_i)\;, \qquad i=2,3\;,$$

$$E_{4r}^{\pm} = \frac{-A_4}{(\chi^{\pm})^2}h_{2E}\cdot\frac{\partial}{\partial r}R_E^{\pm}(r)\Phi_E(\varphi)\exp(-h_{2E}z)\;,$$

$$E_{4\varphi}^{\pm} = \frac{-A_i}{(\chi^{\pm})^2}\frac{h_{2E}}{r}\cdot R_E^{\pm}(r)\frac{\partial}{\partial\varphi}\Phi_E(\varphi)\exp(-h_{2E}z)\,, \qquad \text{(A5)}$$

$$H_{1r}^{\pm} = j\frac{A_1}{(\chi^{\pm})^2}\left[\frac{\omega\varepsilon_0\varepsilon_{1\perp}}{r}\cdot R_E^{\pm}(r)\cdot\frac{\partial}{\partial\varphi}\Phi_E(\varphi)-h\cdot\frac{\partial}{\partial r}R_H^{\pm}(r)\cdot\Phi_H(\varphi)\right]\cdot\cos(hz+\Theta_1)\,,$$

$$H_{1\varphi}^{\pm} = -j\frac{A_1}{(\chi^{\pm})^2}\left[\omega\varepsilon_0\varepsilon_{1\perp}\frac{\partial}{\partial r}R_E^{\pm}(r)\cdot\Phi_E(\varphi)+\frac{h}{r}\cdot R_H^{\pm}(r)\cdot\frac{\partial}{\partial\varphi}\Phi_H(\varphi)\right]\cdot\cos(hz+\Theta_1),$$

$$H_{ir}^{\pm} = j\frac{A_i}{(\chi^{\pm})^2}\frac{\omega\varepsilon_0\varepsilon_i}{r}\cdot R_E^{\pm}(r)\frac{\partial}{\partial\varphi}\Phi_E(\varphi)\cosh(h_{iE}z+\Theta_i)\ ,$$

$$H_{i\varphi}^{\pm} = -j\frac{A_i}{(\chi^{\pm})^2}\omega\varepsilon_0\varepsilon_i\cdot\frac{\partial}{\partial r}R_E^{\pm}(r)\Phi_E(\varphi)\cosh(h_{iE}z+\Theta_i)\ , \qquad i=2,3,$$

$$H_{4r}^{\pm} = j\frac{A_4}{(\chi^{\pm})^2}\frac{\omega\varepsilon_0\varepsilon_2}{r}\cdot R_E^{\pm}(r)\frac{\partial}{\partial\varphi}\Phi_E(\varphi)\exp(-h_{2E}z)\ ,$$

$$H_{4i\varphi}^{\pm} = -j\frac{A_4}{(\chi^{\pm})^2}\omega\varepsilon_0\varepsilon_2\cdot\frac{\partial}{\partial r}R_E^{\pm}(r)\Phi_E(\varphi)\exp(-h_{2E}z)\ ,$$

where $\chi^{+}=\chi_1$ , $\chi^{-}=\chi_2$.

The boundary conditions $E_{1\varphi}^{+}=E_{1\varphi}^{-}, H_{1\varphi}^{+}=H_{1\varphi}^{-}$ for inside (+) and outside (-) field components at $r=a$ within the limits $t+d\le z\le t+d+L$ $(i=1)$ will result in Eq. (1). Assuming the constant $A_3$=1 and neglecting the contributions of $H_{1z}$ in the expressions for $E_{1r}, E_{1\varphi}, H_{1r}, H_{1\varphi}$ to satisfy the boundary conditions at $z=0$, $z=t$, $z=t+d$, $z=t+d+L$ we get the Eq. (2) and the following relations:

$$A_1 = A_2\cdot\frac{\cosh[h_{2E}(t+d)+\Theta_2]}{\varepsilon_{\perp}\cos[h(t+d)+\Theta_1]}\ ,$$

$$A_2 = \frac{\varepsilon_3\cosh(h_{3E}t)}{\varepsilon_2\cosh(h_{2E}t+\Theta_2)}\ ,$$

$$A_4 = A_1\cdot\frac{\varepsilon_{\perp}\cos[h(t+d+L)+\Theta_1]}{\exp[-h_{2E}(t+d+L)]}\ ,$$

$$\Theta_1 = -\mathrm{atan}\left[\frac{\varepsilon_{\perp}h_{2E}}{h}\cdot\tanh[h_{2E}(t+d)+\Theta_2]\right]-h(t+d),$$

$$\Theta_2 = \mathrm{atanh}\left[\frac{\varepsilon_2 h_{3E}}{\varepsilon_3 h_{2E}}\cdot\tanh(h_{3E}t)\right]-h_2 t, \qquad \text{(A6)}$$

$$\Theta_3 = 0.$$

$$I_1 = L+\frac{\sin(2(h(t+d+L)+\Theta_1))-\sin(2(h(t+d)+\Theta_1))}{2h},$$

$$I_2 = d+\frac{\sinh(2(h_{2E}(t+d)+\Theta_2))-\sinh(2(h_{2E}t+\Theta_2))}{2h_{2E}},$$

$$I_3 = t+\frac{\sinh(2h_{3E}t)}{2h_{3E}}, \qquad I_4 = \frac{\exp(-2h_{2E}(t+d+L))}{2h_{2E}}.$$

## APPENDIX B

| SAMPLES INFO | | | SAMPLES PARAMETERS | | | | | | | | | | | | | | |
|---|---|---|---|---|---|---|---|---|---|---|---|---|---|---|---|---|---|
| | | | AVERAGE VALUES | | | | | | VALUES FOR PARTICULAR MODE (9-12) | | | | | | | | |
| Sample/Company Material<br>ε / tanδ x10³ @ 10GHz | Thick-ness, µm | ε (low frequency) | ε<br>$C_\varepsilon$ | ± Δε | tanδ x10³<br>$C_\delta$ | ±Δtanδ x10³ | $R_s$, Ohm | ±Δ $R_s$, Ohm | $f_{diel}$, GHz | $Q_{diel}$ | Error, % | ε | tanδ x10³ | $f_{metal+diel}$, GHz | $Q_{metal+diel}$ | Error, % | $R_s$, Ohm |
| RO3003x0.01"/ Rogers<br>PTFE ceramic<br>3.0 / 1.3 | 255 | 2.92 | 2.85<br>2.43 | 0.03 | 1.3<br>- 1.0 | 0.1 | 0.24 | 0.14 | 31.045<br>33.499<br>35.949<br>38.375 | 9020<br>6833<br>7812<br>8565 | 4.8<br>4.0<br>1.2<br>8.3 | 2.80<br>2.84<br>2.85<br>2.85 | 0.77<br>1.28<br>1.27<br>1.28 | 31.733<br>34.159<br>36.551<br>38.916 | 2806<br>4584<br>5986<br>6694 | 5.4<br>2.9<br>5.8<br>14.1 | 0.38<br>0.24<br>0.22<br>0.28 |
| RO3003x0.02"/ Rogers<br>PTFE ceramic<br>3.0 / 1.3 | 505 | 2.90 | 2.96<br>1.11 | 0.02 | 1.4<br>- 0.49 | 0.2 | 0.17 | 0.02 | 31.670<br>34.099<br>36.490<br>38.854 | 6183<br>6871<br>7526<br>9462 | 2.9<br>0.4<br>1.3<br>3.8 | 2.97<br>2.96<br>2.96<br>2.96 | 1.28<br>1.42<br>1.56<br>1.38 | 36.970<br>39.264<br>0<br>0 | 5345<br>6787<br>0<br>0 | 2.9<br>15.8<br>0<br>0 | 0.17<br>0.18<br>0<br>0 |
| RO3003x0.03"/ Rogers<br>PTFE ceramic<br>3.0 / 1.3 | 760 | 2.94 | 2.93<br>0.640 | 0.02 | 1.7<br>- 0.055 | 0.3 | 0.09 | +0.1<br>-0.04 | 32.035<br>34.418<br>36.759<br>39.080 | 4401<br>6394<br>8054<br>10241 | 13.8<br>4.5<br>4.1<br>1.0 | 2.97<br>2.93<br>2.93<br>2.94 | 2.16<br>1.84<br>1.76<br>1.50 | 32.513<br>34.800<br>37.075<br>39.344 | 3242<br>4219<br>5755<br>9229 | 5.4<br>6.1<br>0.6<br>5.3 | 0.18<br>0.10<br>0.06<br>0.10 |
| NY9220x0.03"/ Nelco<br>Woven PTFE<br>2.2 / 1.3 | 755 | 2.23 | 2.09<br>0.707 | 0.02 | 0.98<br>- 0.16 | 0.14 | 0.110 | 0.016 | 32.478<br>34.768<br>37.053<br>39.327 | 11512<br>13334<br>16623<br>18145 | 0.7<br>7.7<br>3.2<br>3.7 | 2.09<br>2.09<br>2.09<br>2.09 | 0.98<br>1.08<br>1.01<br>0.84 | 32.833<br>35.067<br>37.297<br>39.531 | 7720<br>11570<br>14484<br>16103 | 10.9<br>1.4<br>7.1<br>1.7 | 0.12<br>0.11<br>0.10<br>0.11 |
| NY9220x0.01"/ Nelco<br>Woven PTFE<br>2.2 / 1.3 | 245 | 2.25 | 2.21<br>2.97 | 0.02 | 0.86<br>- 0.87 | 0.26 | 0.25 | 0.03 | 31.400<br>33.854<br>36.274<br>38.667 | 10585<br>10733<br>10290<br>12360 | 5.4<br>0.8<br>2.0<br>2.7 | 2.22<br>2.21<br>2.20<br>2.21 | 0.69<br>0.83<br>1.06<br>0.95 | 32.149<br>34.519<br>36.860<br>39.181 | 6064<br>7902<br>9863<br>11470 | 8.1<br>1.3<br>5.5<br>1.8 | 0.230<br>0.244<br>0.256<br>0.259 |
| GIG M/L-2/isola | 75 | 3.85 | 2.90<br>9.74 | 0.02 | 7.8<br>- 3.8 | 0.7 | 0.97 | | 29.835<br>32.314<br>34.784<br>37.243 | 1798<br>1829<br>2060<br>1956 | 1.5<br>2.5<br>1.7<br>5.2 | 2.89<br>2.90<br>2.90<br>2.92 | 7.91<br>8.10<br>7.51<br>8.32 | 30.268<br>35.321<br>0<br>0 | 329<br>71<br>0<br>0 | 9.4<br>49.1<br>0<br>0 | 1.24<br>0.73<br>0<br>0 |
| NH9348ST0203CHCH/<br>Neltec<br>PTFE/glass/ceramic<br>3.48 / 3 | 205 | 3.60 | 3.32<br>2.94 | 0.03 | 3.1<br>- 1.6 | 1.2 | 0.36 | | 30.512<br>35.499<br>37.949<br>0 | 2924<br>3270<br>601<br>0 | 2.7<br>0.8<br>11.1<br>0 | 3.32<br>3.30<br>3.33<br>0 | 2.95<br>3.27<br>21.3<br>0 | 31.191<br>33.929<br>36.111<br>0 | 1308<br>1390<br>2566<br>0 | 2.3<br>11.8<br>4.1<br>0 | 0.25<br>0.15<br>0.46<br>0 |
| NY9208ST0762CHCH/<br>Neltec<br>PTFE woven<br>2.08 / 0.6 | 800 | 2.20 | 1.96<br>0.585 | 0.02 | 0.49<br>- 0.11 | 0.15 | 0.12 | 0.06 | 32.573<br>34.846<br>37.117<br>39.379 | 18632<br>20234<br>23638<br>25726 | 2.5<br>10.3<br>10.4<br>4.7 | 1.96<br>1.96<br>1.95<br>1.96 | 0.49<br>0.57<br>0.52<br>0.41 | 32.898<br>35.121<br>37.341<br>39.568 | 12081<br>16826<br>21609<br>23930 | 6.2<br>3.0<br>3.7<br>5.7 | 0.12<br>0.11<br>0.17<br>0.11 |
| tly5a0200/Taconic<br>PTFE/glass<br>2.17 / 0.4 | 525 | 1.80 | 2.12<br>1.22 | 0.02 | 0.94<br>- 0.32 | 0.05 | 0.09 | 0.04 | 32.219<br>34.546<br>36.872<br>39.181 | 10355<br>12691<br>14400<br>16618 | 3.5<br>2.1<br>2.1<br>2.0 | 2.11<br>2.13<br>2.13<br>2.13 | 0.93<br>0.93<br>0.97<br>0.92 | 32.700<br>34.966<br>37.221<br>39.474 | 6228<br>9766<br>13156<br>13793 | 2.5<br>0.7<br>4.7<br>2.5 | 0.06<br>0.11<br>0.08<br>0.09 |
| teflon395 | 395 | 2.00 | 2.03<br>1.76 | 0.03 | 0.19<br>- 0.47 | 0.05 | 0 | 0 | 32.005<br>34.403<br>36.755<br>39.086 | 22823<br>27812<br>28090<br>30672 | 3.6<br>1.0<br>0.7<br>1.5 | 2.06<br>2.01<br>2.00<br>1.99 | 0.20<br>0.18<br>0.20<br>0.15 | 0 | 0 | 0 | 0 |
| Cu-coated quartz 2 | 1000 | 3.80 | 3.88<br>0.362 | 0.03 | 0.63<br>- 0.18 | 0.07 | 0 | 0 | 31.791<br>34.201<br>36.572<br>38.916 | 10118<br>12896<br>14161<br>14772 | 100.0<br>100.0<br>99.4<br>9.8 | 3.87<br>3.86<br>3.87<br>3.90 | 0.60<br>0.57<br>0.61<br>0.66 | 0 | 0 | 0 | 0 |
| Rogers 6002<br>2.94 / 1.2 | 395 | 2.94 | 2.83<br>1.56 | 0.02 | 1.2<br>- 0.67 | 0.1 | 0.13 | +0.15<br>-0.08 | 31.502<br>33.949<br>36.360<br>38.742 | 6507<br>7768<br>8765<br>10371 | 1.7<br>1.1<br>1.0<br>5.2 | 2.84<br>2.82<br>2.82<br>2.83 | 1.19<br>1.19<br>1.24<br>1.15 | 34.546<br>36.877<br>39.190<br>0 | 6415<br>7488<br>7800<br>0 | 10.5<br>8.9<br>5.1<br>0 | 0.06<br>0.12<br>0.21<br>0 |
| Rogers 3003<br>PTFE ceramic<br>3.0 / 1.3 | 245 | 3.15 | 2.89<br>2.51 | 0.03 | 1.2<br>- 1.1 | 0.2 | 0.23 | 0.04 | 30.925<br>33.433<br>35.885<br>38.315 | 6768<br>8093<br>8001<br>8283 | 1.9<br>3.4<br>1.7<br>4.4 | 2.93<br>2.89<br>2.89<br>2.90 | 1.10<br>1.04<br>1.21<br>1.31 | 31.682<br>34.116<br>36.513<br>38.882 | 3617<br>4997<br>6197<br>6988 | 8.4<br>6.1<br>6.1<br>0.9 | 0.22<br>0.19<br>0.20<br>0.24 |
| Rogers 5880<br>PTFE/glass/fiber<br>2.2 / 0.9 | 120 | 2.34 | 2.16<br>6.14 | 0.02 | 0.74<br>- 1.8 | 0.25 | 0.21 | 0.06 | 30.632<br>33.115<br>35.608<br>38.049 | 11703<br>12756<br>10266<br>13209 | 2.8<br>1.5<br>4.4<br>4.3 | 2.17<br>2.18<br>2.14<br>2.16 | 0.63<br>0.65<br>0.98<br>0.73 | 31.389<br>33.848<br>36.274<br>38.673 | 4114<br>4579<br>6067<br>6463 | 3.9<br>5.7<br>1.4<br>3.5 | 0.175<br>0.219<br>0.197<br>0.242 |
| Sheldahl G2200x2mil | 50 | 1.4 | 2.99<br>13.3 | 0.02 | 12.9<br>- 5.8 | 0.9 | 0.13 | 0.02 | 29.561<br>32.024<br>34.481<br>36.933 | 1442<br>1513<br>1547<br>1590 | 1.2<br>4.4<br>1.1<br>4.5 | 2.98<br>2.98<br>2.99<br>3.01 | 13.1<br>12.7<br>12.8<br>12.8 | 30.050<br>32.538<br>35.016<br>37.479 | 922<br>698<br>932<br>534 | 3.3<br>1.8<br>1<br>46.4 | 0.01<br>0.14<br>0.12<br>0.12 |
| Sheldahl Comclad XFx10mil | 280 | 2.27 | 2.24<br>2.48 | 0.02 | 0.44<br>- 0.77 | 0.12 | 0.07 | +0.05<br>-0.02 | 31.518<br>33.962<br>36.370<br>38.753 | 20103<br>15778<br>16270<br>18707 | 5.1<br>25.6<br>22.0<br>29.1 | 2.26<br>2.24<br>2.24<br>2.24 | 0.24<br>0.49<br>0.55<br>0.47 | 32.243<br>34.598<br>36.924<br>39.235 | 10044<br>17829<br>20839<br>22934 | 4.3<br>2.9<br>3.3<br>3.0 | 0.10<br>0.07<br>0.05<br>0.05 |
| tlc32-031 | 805 | 2.70 | 2.92<br>0.576 | 0.03 | 5.3<br>- 0.21 | 0.5 | 0.21 | | 32.102<br>34.455<br>36.799<br>39.12 | 1116<br>2641<br>3202<br>4015 | 2.1<br>3<br>6.1<br>3.2 | 2.92<br>2.93<br>2.90<br>2.89 | 9.84<br>5.15<br>5.38<br>5.24 | 37.073<br>39.348<br>0<br>0 | 3511<br>3850<br>0<br>0 | 4.6<br>0.5<br>0<br>0 | 0.39<br>0.03<br>0<br>0 |
| rdx1100060 | 155 | 3.72 | 3.07<br>3.92 | 0.03 | 4.2<br>- 1.9 | 0.3 | 0.13 | | 30.365<br>32.857<br>35.425<br>37.816 | 2208<br>2508<br>5531<br>2998 | 2.3<br>3.6<br>0.8<br>3.8 | 3.07<br>3.08<br>2.91<br>3.05 | 4.32<br>4.14<br>1.92<br>4.12 | 35.977<br>0<br>0<br>0 | 2054<br>0<br>0<br>0 | 2.5<br>0<br>0<br>0 | 0.13<br>0<br>0<br>0 |